\documentclass[aps,pra,numerical,reprint,showpacs]{revtex4-1}  
\usepackage[dvips]{graphicx}%
\usepackage{bm}%
\usepackage{multirow}
\usepackage{amsmath}
\usepackage[normalem]{ulem} 
\usepackage[colorlinks=true,linkcolor=blue,citecolor=blue ,urlcolor=blue]{hyperref}

\begin{document}

\newcommand{\new}[1]{\textcolor[rgb]{0,0.5,0.3}{#1}}
\newcommand{\old}[1]{\textcolor[rgb]{1,0,0}{\sout{#1}}}


\title{Polychromatic phase diagram for $n$-level atoms interacting with $\ell$ modes of electromagnetic field}
\author{S. Cordero} 
\email{sergio.cordero@nucleares.unam.mx}
\author{E. Nahmad--Achar}
%
\author{R. L\'opez--Pe\~na}
%
\author{O. Casta\~nos}
%

%
\affiliation{%
Instituto de Ciencias Nucleares, Universidad Nacional Aut\'onoma de M\'exico, Apartado Postal 70-543, 04510 M\'exico DF,   Mexico }

\date{\today}

\begin{abstract}
A system of $N_a$ atoms of $n$-levels interacting dipolarly with $\ell$ modes of electromagnetic field is considered. The energy surface of the system is constructed from the direct product of the coherent states of U$(n)$ in the totally symmetric representation for the matter times the $\ell$ coherent states of the electromagnetic field. A variational analysis shows that the collective region is divided into $\ell$ zones, inside each of which only one mode of the electromagnetic field contributes to the ground state. In consequence, the polychromatic phase diagram for the ground state naturally divides itself into monochromatic regions. For the case of $3$-level atoms in the $\Xi$-configuration in the presence of $2$ modes, the variational calculation is compared with the exact quantum solution showing that both  are in agreement. 
\end{abstract}

%

\maketitle

\section{Introduction}
Research in the interaction between matter and electromagnetic radiation has proved very fruitful no less for its interest in quantum information theory, a field emerging from quantum optics and information science. In order to study the interaction between matter and radiation a simple but non trivial model,  which considers $N_a$  identical atoms of $2$-levels under dipolar interaction with one-mode of an electromagnetic field was proposed by Dicke~\cite{dicke54}. The Tavis-Cummings model~\cite{tavis68,tavis69} is a simplification which considers the rotating wave approximation (RWA), eliminating the terms in the Hamiltonian that do not preserve the number of excitations in the system. The particular case of a single $2$-level atom with the RWA approximation is known as the Jaynes-Cummings model~\cite{jaynes63} and has analytical solution. For the case without the RWA approximation, Hepp and Lieb found  in the early seventies that the system suffers a transition from the normal to the superradian phase~\cite{hepp73, hepp73b}

Whereas $n$-levels atoms are only an approximation to real atoms, the design and construction of artificial quantum structures allows one to refer to the so-called artificial atoms~\cite{kastner93,astafiev10,buluta11} that possess a finite number of levels. Hence it is interesting to consider $n$-level systems without loss of generality.

A first generalisation of the models involving $2$-level atoms is the consideration of atoms of three or more levels~\cite{yoo85,abdel-wahab07,abdel-wahab08} which, as is well known, possess different  arrangements: the $\Xi$-, $V$- and $\Lambda$-configurations for $3$-level atoms, and  for $4$-level atoms the configurations $\Xi$, $\lambda$, ${\rm y}$, $\diamondsuit$, $N$, \dots. Clearly, systems of three or more levels permit to consider dipolar interactions with one, two or more modes of electromagnetic field. In fact, depending on the atomic configuration, there is a maximum number of different modes that may produce atomic transitions via the dipolar interaction. For $3$-level atoms, the interaction with one-mode of the electromagnetic field has been extensively studied~\cite{yoo85,civitarese1,abdel-wahab07,cordero1,cordero2}. More recently, treatments including the
diamagnetic term show that, in spite of the
Thomas-Reiche-Kuhn rule for multilevel atoms, there is a broad range of
physical interaction parameters for which it is possible to have a
superradiant transition~\cite{baksic13}; and under a degenerate parametric amplification process and with an external laser driving field the system exhibits the effects of spontaneous emission of atoms and cavity loss in the phase diagram~\cite{yu14}. Two-modes of electromagnetic field have also been considered in~\cite{yoo85,ting06,civitarese2}. In particular, the phase diagram in the $\Lambda$-configuration was shown to be constituted by a normal and {\it two} superradiant phases~\cite{hayn11,hayn12}.

In this work we study the properties of the variational ground state of $N_a$ atoms of $n$-levels interacting dipolarly with $\ell$ modes of electromagnetic field, where each mode promotes transitions only between two given atomic levels and where $\ell$ is the maximum number of modes that are permitted by the dipolar interaction. The variational ground state is evaluated by considering as test function the direct product of Heisenberg-Weyl coherent states for the field and U$(n)$ coherent states for the matter.  The variational energy surface is evaluated as the expectation value of the Hamiltonian for this test state. We find in general that  critical points both at zero and at  infinity (for the matter variables) provide minimum values of the energy, and that taking particular values of them (at zero or at infinity)  reduces the system to $(n-1)$-levels, which may again be reduced to an $(n-2)$-level system and so on,  through this iterative process in the end one has reduced systems of the well-known {\em Dicke model}, which in the variational method can be solved analytically. The method is exemplified for the cases of $3$- and $4$-level atoms.  General expressions of the critical points that provide the minimum energy surface are obtained as functions of the dipolar intensities, and from them the corresponding normal and collective (superradiant) regions are determined for the case of $n$-level atoms.  

We show that the collective region is divided into $\ell$ different zones where in each zone  only one-mode of electromagnetic field contributes to the ground state, i.e., the polychromatic phase diagram is divided into monochromatic regions, and crossing from one to the other represents a first order phase transition. These transitions manifest themselves as discontinuities in the atomic populations. On the other hand, crossings from the normal region to any of the collective monochromatic region give place to first or second order phase transitions; those of second order  correspond to bifurcations, while those of first order form  Maxwell sets~\cite{gilmore93}.    This analysis leads to a universal relationship between expectation value of the number of photons in the mode $\Omega_{jk}$, $\langle {\bm \nu}_{jk}\rangle$, and the quadratic fluctuations of the  number of atoms  in the level $j$, $(\Delta {\bm A}_{jj})^2$. We propose this relation  as an experimental criterion to detect the transition between the normal and superradiant regimes.

This paper is organized as follows: 
Section~\ref{model} describes the Hamiltonian for a system of $N_a$ atoms of $n$ levels under dipolar interaction with $\ell$ modes of electromagnetic field, where only one mode promotes transitions between two given atomic levels. 
In section~\ref{energy.surface}, the variational ground state (test function) is defined for the field and matter contributions. The energy surface is calculated and the method (which involves an iterative reduction of the system) to find the critical points that minimize the energy is discussed. 
In section~\ref{s.3l} the method is exemplified for the case of $3$-level atoms interacting dipolarly with $2$-modes of electromagnetic field is considered for the three atomic configurations. The minimum energy surface, the order of transitions, and the statistical properties of the variational ground state are established. 
In section~\ref{s.4l} the critical points of the energy surface for the case of $4$-level atoms in dipolar interaction with $\ell$ modes ($\ell=3,4$) are obtained by means of the iterative procedure.    
Section~\ref{s.gen} generalizes the expressions to the case of $n$-level atoms interacting dipolarly with $\ell$ modes. Also, we discuss in general the quantum phase diagram, the order of the transitions, the expectation values of field and matter observables and the corresponding results when the RWA approximation is considered.  Finally in section~\ref{s.quantum} the numerical exact quantum solution is given for $3$-level atoms in the $\Xi$-configuration for $N_a=1$ and $N_a=2$ particles.
In section~\ref{conclusions} we give some concluding remarks.

\section{Model}\label{model}

We assume that any given pair of atomic levels $\omega_j$ and $\omega_k$ can be connected by at most one radiation mode denoted by the photon annhilation $\bm{a}_{jk}$ and creation $\bm{a}^{\dagger}_{jk}$ operators. The Hamiltonian describing a system of $N_a$ atoms of $n$-levels on dipolar interaction with $\ell$ modes of electromagnetic field is then written as~\cite{cordero2}
\begin{equation}\label{eq.H}
\bm{H} = \bm{H}_D + \bm{H}_{int},
\end{equation}
with a diagonal contribution
\begin{equation}\label{eq.HD}
\bm{H}_D = \sum_{j<k}^{n} \Omega_{jk}\, \bm{a}_{jk}^\dag\, \bm{a}_{jk} + \sum_{j=1}^{n} \omega_j \, \bm{A}_{jj}\, .
\end{equation}
We have taken $\hbar=1$, and the collective atomic operators  $\bm{A}_{kj}$ obey a U$(n)$ algebra satisfying the commutation relations
\begin{equation}
\left[\bm{A}_{lm},\bm{A}_{kj}\right] = \delta_{mk}\,\bm{A}_{lj}-\delta_{jl}\,\bm{A}_{km}\, ,\label{eq.commAij} 
\end{equation}
with the first order Casimir operator determining the total number of atoms $N_{a}$ given by
\begin{equation}
\sum_{k=1}^n \bm{A}_{kk} = N_a, \label{eq.Na}
\end{equation} 
which is conserved. We denote field frequencies by $\Omega_{jk}$ and assume that the atomic frequencies satisfy $\omega_1<\omega_2<\cdots<\omega_{n}$.

The interaction $\bm{H}_{int}$ between the atoms and radiation field involves contributions of the form $(\bm{A}_{jk}+\bm{A}_{kj})(\bm{a}_{jk} + \bm{a}_{jk}^\dag)$ representing the fact that  a transition between the levels $j$ and $k$ is only promoted by the  mode $\Omega_{jk}$. Thus we may write
\begin{equation}\label{eq.Hint}
\bm{H}_{int} =- \frac{1}{\sqrt{N_a}} \sum_{j<k}^{n} \mu_{j k} \left(\bm{A}_{jk}+\bm{A}_{kj}\right)\left(\bm{a}_{jk} + \bm{a}_{jk}^\dag\right)
\end{equation}
%
%
where $\mu_{jk}$ denote the dipolar intensities. 

The maximum number of dipolar interaction strengths of an $n$-level system is $\ell_{max} = n(n-1)/2- (n-2)$; of course the real number $\ell$ of modes present in the system will be less than or equal to $\ell_{max}$ and it depends of the considered atomic configuration with its allowed dipolar transitions.   
%

\section{Energy surface}\label{energy.surface}

In order to find the minimum energy surface, we propose as a variational test function a coherent state containing contributions of both field and matter.

The coherent field contribution is well known, since each electromagnetic mode $\Omega_{jk}$ is described by the Heisenberg-Weyl state $|\alpha_{jk}\rangle$ obeying $\bm{a}_{jk}\,|\alpha_{jk}\rangle = \alpha_{jk}\,|\alpha_{jk}\rangle$. The dimension of vector $\vec{\alpha}=(\alpha_{jk})$, $j<k=1,\,2,\,\dots,\, n$ depends on the number of allowed dipolar transitions in the given atomic configuration.
The total coherent electromagnetic field contribution is  then given by the direct product of these states, {\em i.e.},
\begin{equation}
|\vec{\alpha}\rangle = \bigotimes_{\substack{j<k}}^{n} |\alpha_{jk}\rangle\,.
\end{equation}

The form of the coherent matter contribution depends in general on the representation of the atomic operators; since we are considering identical particles one may use the boson representation $\bm{A}_{jk} = \bm{b}^\dag_j\,\bm{b}_k$, where  $[\bm{b}_k,\bm{b}_l^\dag]=\delta_{kl}$ with $\delta_{kl}$ the Kronecker symbol. The coherent state for the matter contribution that preserves the number $N_a$ of atoms is then written as
\begin{equation}\label{eq.coh.matter}
|\vec{\gamma}\rangle =\frac{1}{\sqrt{N_a!}}\, \left[\bm{\Gamma}^\dag\right]^{N_a}|0\rangle_{M},
\end{equation} 
where the $\bm{\Gamma}^\dag$ operator is defined as
\begin{equation}
\bm{\Gamma}^\dag = \frac{\gamma_1\,\bm{b}_1^\dag + \gamma_2\,\bm{b}_2^\dag + \cdots + \gamma_n\,\bm{b}_n^\dag}{\left(|\gamma_1|^2 + |\gamma_2|^2 +\cdots + |\gamma_n|^2\right)^{1/2}},
\end{equation}
and where $\vec{\gamma} = (\gamma_1,\,\gamma_2,\,\dots,\,\gamma_n)$. Here $|0\rangle_{M}$ represents the vacuum state in the Fock basis. From the relationship
\begin{eqnarray}
[\bm{b}_k,\bm{\Gamma}^\dag] &=& \frac{\gamma_k}{\left(\sum_{l=1}^n |\gamma_l|^2\right)^{1/2}},\quad k=1,\,2,\,\dots,\,n,\nonumber
\end{eqnarray}
one finds the identity $[\bm{\Gamma},\bm{\Gamma}^\dag]=1$. It is worth noticing that this identity shows that $\bm{\Gamma}$ and $\bm{\Gamma}^\dag$ obey a bosonic algebra, hence the coherent state defined in Eq.~(\ref{eq.coh.matter}) is normalized. One may obtain immediately
\begin{eqnarray}
[\bm{b}_k,\left[\bm{\Gamma}^\dag\right]^{N_a}] &=& \frac{N_a\,\gamma_k}{\left(\sum_{l=1}^n |\gamma_l|^2\right)^{1/2}}\left[\bm{\Gamma}^\dag\right]^{N_a-1}.
\end{eqnarray}

Without loss of generality, one may diminish the number of variables by choosing 
$\gamma_1=1$. In fact, this assumption is strictly equivalent to eliminate a global phase in the matter coherent state (\ref{eq.coh.matter}) and renormalize the other values of $\gamma_k$. 

In order to find the expectation value of the Hamiltonian, the matrix elements of the matter operators are required. These read
\begin{equation}\label{eq.Ajk}
\langle\vec{\gamma}|\bm{A}_{jk}|\vec{\gamma}\rangle = \frac{N_a\,\gamma_j^*\,\gamma_k}{1+ \sum_{l=2}^n |\gamma_l|^2},
\end{equation}
where we have used explicitly in the denominator the fact that $|\gamma_1|^2=1$.

Finally, the test state that possesses the contribution of matter and field is written as
\begin{eqnarray}\label{eq.test.state}
|\vec{\alpha},\,\vec{\gamma}\rangle = |\vec{\alpha}\rangle\,\otimes\,|\vec{\gamma}\rangle.
\end{eqnarray}
Using this as a variational state one finds the energy surface
\begin{eqnarray}\label{eq.E.test}
{\cal E}&=& \langle \vec{\alpha},\vec{\gamma}| \bm{H}|\vec{\alpha},\vec{\gamma}\rangle
\nonumber\\
&=& \sum_{j<k}^n \Omega_{jk}\, {R}_{jk}^2 +  N_a \sum_{j=1}^n\omega_j \frac{\varrho_j^2}{1+R^2_0}  \nonumber \\
&& - 4 \sqrt{N_a} \sum_{j<k}^n \mu_{jk}\,{R}_{jk}\frac{\varrho_{j}\,\varrho_{k}\, \cos(\theta_{jk})\,\cos(\phi_{k}-\phi_{j})}{1+R^2_0},\nonumber\\
\end{eqnarray}
where we have written $\alpha_{jk}= {R}_{jk}\, e^{i\,\theta_{jk}}$ and $\gamma_k = \varrho_k\, e^{i\,\phi_k}$ for $j,\,k=1,\,2,\,\dots,\,n$,  with $\varrho_1=1$ and $\phi_1=0$. Additionally, we have defined $R^2_0 = \sum^n_{j=2}\,  \varrho^2_{j}$.


An estimation of the ground state energy is obtained by minimising the expression (\ref{eq.E.test}). Differentiating with respect to the phases $\theta_{jk}$ and $\phi_{k}$ one finds critical points at
\begin{equation}\label{eq.phases.c}
\theta^{c}_{jk} = 0,\, \pi, \qquad \phi^{c}_{k }-\phi^{c}_{j } = 0,\, \pi \, ,
\end{equation}
and by simple inspection the condition for a minimum reads 
\begin{displaymath}
\mu_{jk}\,\cos\left(\theta^{c}_{jk}\right)\,\cos\left(\phi^{c}_{k }-\phi^{c}_{j}\right)>0.
\end{displaymath}
Assuming $\mu_{jk}>0$ the minimum is obtained at  $\theta^{c}_{jk} =  \phi^{c}_{k }-\phi^{c}_{j } = 0,\, \pi$, and replacing it into Eq. (\ref{eq.E.test}) the minimum energy surface is reduced to
\begin{eqnarray}\label{eq.E.test1}
{\cal E} &=& \sum_{j<k}^n \Omega_{jk}\, {R}_{jk}^2 + N_a \sum_{j=1}^n\omega_j \frac{\varrho_j^2}{1+R^2_0}  \nonumber \\
&& - 4 \sqrt{N_a} \sum_{j<k}^n \mu_{jk}\,{R}_{jk}\frac{\varrho_{j}\,\varrho_{k}}{1+R^2_0}.
\end{eqnarray}

Differentiating the energy surface with respect to the variables ${R}_{jk}$ (related to the expectation value of photon number in the $jk$ mode) one finds critical points at
\begin{equation}\label{eq.rho.c}
{R}^{c}_{jk} = 2\,\mu_{j k}\,\frac{\sqrt{N_a}}{\Omega_{jk}} \,  
\frac{\varrho^{c}_{j}\,\varrho^{c}_{k}}{1+R^{c \, 2}_{0}} \equiv \sqrt{N_a}\, r^c_{jk} ,
\end{equation}
where $\varrho^c_{k}$ stands for the critical value of $\varrho_k$ ({\it vide infra}). Also, since the energy surface is an extensive quantity it makes sense to normalize it with respect to the number of particles $E={\cal E}/N_a$. Additionally, without loss of generality we choose $\omega_1=0$, and hence the energy surface per particle maybe written as
\begin{eqnarray}\label{eq.E.test2}
E &=& \sum_{j<k}^n \Omega_{jk}\, r^{c \, 2}_{jk} +  \sum_{j=2}^n\omega_j \frac{\varrho_j^2}{1+R^2_0}  \nonumber \\
&& - 4  \sum_{j<k}^n \mu_{jk}\,r^{c}_{jk}\frac{\varrho_{j}\,\varrho_{k}}{1+R^2_0}.
\end{eqnarray}

Finally, the critical points for the variables $\varrho_k$ are obtained by solving the system of coupled equations  
\begin{eqnarray}
\label{eq.dErhos}
\frac{\partial E}{\partial \varrho_{j}}\bigg|_{\{\varrho^{c}\}} &=& \frac{\varrho_j} {1 + R^2_0} \Bigg( \omega_j - \sum_{j'}\frac{  \omega_{j'} \, 
\varrho^2_{j'}}{1 + R^2_0} \nonumber \\
&-&4 \sum^n_{k=j+1} \frac{\mu^2_{j k} \varrho^2_{k}}{\Omega_{j k} (1 + R^2_0)}  \nonumber \\
&-& 4 \sum^{j-1}_{k=1} \frac{\mu^2_{k j} \varrho^2_{k}}{\Omega_{k j} (1 + R^2_0)} \nonumber \\
&+&  8 \sum^n_{j'<k'} \frac{\mu^2_{j' k'} \varrho^2_{j'} \varrho^2_{k'}}{\Omega_{j' k'} (1 + R^2_0)^2} \Bigg)\Bigg|_{\{\varrho^{c}\}} =0 \, ,
\end{eqnarray}
for $j=2,\,3,\,\dots,\,n$, where $\{\varrho^{c}\}=\{\varrho^{c}_{2},\,\dots,\,\varrho^c_{n}\}$ represents the set of critical points. At these points the energy surface becomes, using (\ref{eq.rho.c}), 
\begin{eqnarray}
\label{E-critica}
E &=&  \sum_{j=2}^n\omega_j \frac{\varrho^{c \, 2}_j}{1+R^{c \, 2}_{0}} \nonumber \\
&-& 4  \sum_{1 \leq j<k}^n \, \frac{\mu^2_{jk}}{\Omega_{jk}} 
\left(\frac{\varrho^c_{j} \,\varrho^c_{k}}{1+R^{c \, 2}_{0}}\right)^2 \, .
\end{eqnarray}

Notice that for any value of the dipolar strengths, $\varrho^c_j=0$ and $\varrho^c_{j}=\infty$ are critical points. 
In particular, when $\varrho^c_j=0$ for all $j$ the energy surface takes the value $E=0$.  On the other hand, when a particular $\varrho^c_s=0$ for $s=2,\, 3,\,\dots,\,n$,  all the dipolar coupling interactions terms related to this level are zero. Thus it is straightforward that the atomic part reduces effectively to an $(n-1)$-level system, whose critical points must be analyzed. Hence,  in this reduction scheme, a given atomic configuration will usually lead to different $2$-level reduced set; one then selects the reduced set with minimal energy in order to study the variational ground state of the original system

If for instance, we consider the limit $\varrho_{2}\to\infty$, it is straightforward that the energy surface takes the value
	\begin{equation}
		\lim_{\varrho_2 \to\infty}{E}=\omega_{2} \, ,
	\end{equation}
which is greater than zero. The interaction terms, which give a negative contribution to the energy surface, are different from zero (in this limit) if and only if we assume that the critical points satisfy
\begin{equation}
\varrho^c_j \to \eta_j \, \varrho^c_2  \, , \qquad \varrho^c_2 \to \infty \, ,
\end{equation}
for $j=2, 3, \cdots, n$. From the expression of the energy surface one gets 
\begin{eqnarray}
\label{E-infinity}
E_\infty &=&  \sum_{j=2}^n\omega_j \frac{\eta^{\, 2}_j}{1+\sum_{k=3}^n \eta^{ 2}_k} \nonumber \\
&-& 4  \sum_{2 \leq j<k}^n \, \frac{\mu^2_{jk}}{\Omega_{jk}} 
\left(\frac{\eta_{j} \,\eta_{k}}{1+\sum_{\ell=3}^n \eta ^{ 2}_{\ell}}\right)^2 \, ,
\end{eqnarray}
where $\eta_2\equiv1$ and we have used the relations
\begin{eqnarray}
r^c_{ij} &\to& 2 \frac{\mu_{ij}}{\Omega_{ij}} \, \frac{\eta_{i} \,\eta_{j}}{1+\sum_{k=3}^n \eta_{k}^2} \, , \nonumber \\
\frac{\varrho_{i}^c \,\varrho_{j}^c}{1+\sum_{k=2}^n {\varrho_{k}^c}^2} &\to& \frac{\eta_{i} \,\eta_{j}}{1+\sum_{k=3}^n \eta_{k}^2} \, . 
\end{eqnarray}
Comparing (\ref{E-infinity}) with (\ref{E-critica}) it is clear that we have obtained an equivalent system with $(n-1)$-levels, where the atomic variables $\varrho_j$ are replaced by the new variables $\eta_j$ which upon finding their critical values lead to an equivalent algebraic system of $(n-2)$-levels and so on until we reach a  two-level system as described above, with one radiation mode whose properties, including the complete structure of the phase diagram, have been studied extensively~\cite{castanos11,castanos11b,nahmad-achar13,nahmad-achar15}.  

The complete solution of  (\ref{eq.dErhos}) in general requires numerical techniques. However, by using the physical considerations mentioned above, one may find the minimum energy surface and hence the variational ground state. This new methodology will be exemplified by considering systems of $3$- and $4$-level atoms in the next sections and the results are generalised for $n$-level atoms.


\begin{figure}
\begin{center}
\includegraphics[width=0.45\linewidth]{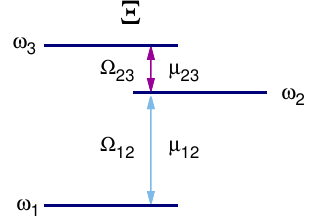}
\includegraphics[width=0.45\linewidth]{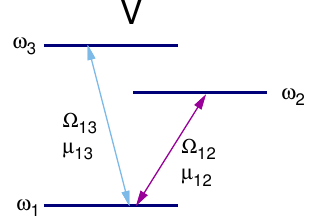}\\[8mm]
\includegraphics[width=0.45\linewidth]{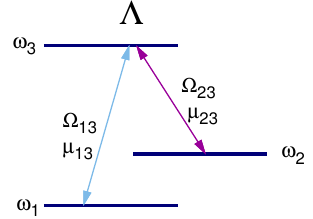}
\caption{Configurations of a $3$-level atom interacting with two-modes of electromagnetic field. For the $\Xi$-configuration $\mu_{13}=0$, for the $\Lambda$-configuration $\mu_{12}=0$, and for the $V$-configuration $\mu_{23}=0$. The mode of the electromagnetic field $\Omega_{jk}$ for each transition is indicated.}\label{conf3n}
\end{center}
\end{figure}

\section{$3$-level atoms}\label{s.3l}

Taking $n=3$ in the expression for the variational energy surface~(\ref{E-critica}), one has
\begin{eqnarray}
E &=& \frac{1}{1 + R_0^2} \Big(\omega_2 \, \varrho^{ c \, 2}_2 + \omega_3 \, \varrho^{ c \, 2}_3 \Big) -  \frac{4}{(1 + R_0^2)^2} \frac{\mu^2_{23}  \, \varrho^{ c \, 2}_2\, \varrho^{ c \, 2}_3 }{\Omega_{23}}  \nonumber \\
 &-& \frac{4}{(1 + R_0^2)^2} \frac{\mu^2_{12} \, \varrho^{ c \, 2}_2}{\Omega_{12}} - \frac{4}{(1 + R_0^2)^2} \frac{\mu^2_{13} \, \varrho^{ c \, 2}_3}{\Omega_{13}} \, ,  
\end{eqnarray}
where without loss of generality we have taken $\omega_1=0$.  In figure~\ref{conf3n} the three different atomic configurations are shown; also the value $\mu_{jk}=0$ that defines each configuration is given.  From the analysis of the previous section
$(\varrho^{ c}_2,\varrho^c_3) =(0,0)$ is a critical point which yields an energy surface equal to zero, $E_{{\rm N}}=0$ which defines the {\em normal} phase.

By considering $\varrho^c_3 =0$ and $\varrho^c_2 \neq 0$ , the energy surface takes the form 
\begin{eqnarray}
E &=& \frac{1}{1 + \varrho^{c \, 2}_2} \omega_2 \, \varrho^{ c \, 2}_2  -  \frac{4}{(1 +  \varrho^{c \, 2}_2)^2} \frac{\mu^2_{12} \, \varrho^{ c \, 2}_2}{\Omega_{12}}  \, ,  
\end{eqnarray}
whose critical points and corresponding minimum can be obtained exactly as
\begin{equation}
\label{eq.varrho.k2}
\varrho^{c}_2 = \sqrt{\frac{4 \mu^2_{12} - \Omega_{12} \, \omega_2}{4 \mu^2_{12} + \Omega_{12} \, \omega_2}} \, , \quad   \, E_{12} = -\frac{(- 4 \, \mu^2_{12} +  \Omega_{12} \, \omega_2)^2}{16 \mu^2_{12} \, \Omega_{12}} \, ,
\end{equation}
where the solution exists only for values of the dipolar strength $ 4 \, \mu^2_{12} \geq  \Omega_{12} \, \omega_2$. The equality in this last expression establishes the border between the normal and a superradiant monochromatic region with photonic mode $\Omega_{12}$.

By considering $\varrho^c_2 =0$ and $\varrho^c_3 \neq 0$, the energy surface takes the form 
\begin{eqnarray}
E &=& \frac{1}{1 + \varrho^{ c \, 2}_3}  \omega_3 \, \varrho^{ c \, 2}_3  - \frac{4}{(1 + \varrho^{ c \, 2}_3)^2} \frac{\mu^2_{13} \, \varrho^{ c \, 2}_3}{\Omega_{13}} \, .   
\end{eqnarray}
Its critical point has the same expression as in the previous case by replacing $\omega_2 \to \omega_3$, $\mu_{12} \to \mu_{13}$, and $\Omega_{12} \to \Omega_{13}$, that is,
\begin{equation}
\label{eq.varrho.k3}
\varrho^{c}_3 = \sqrt{\frac{4 \mu^2_{13} - \Omega_{13} \, \omega_3}{4 \mu^2_{13} + \Omega_{13} \, \omega_3}} \, , \quad   \, E_{13} = -\frac{(- 4 \, \mu^2_{13} +  \Omega_{13} \, \omega_3)^2}{16 \mu^2_{13} \, \Omega_{13}} \, .
\end{equation}
This solution exists only for values of the dipolar strength $ 4 \, \mu^2_{13} \geq  \Omega_{13} \, \omega_3$, where the equality fixes the border between the normal and the superradiant monochromatic region with photonic mode $\Omega_{13}$.

By making the replacement $\varrho_3^c\to\eta_3 \,\varrho_2^c$  into the expression for the energy surface, and taking the limit $\varrho_2^c\to\infty$, one gets
\begin{eqnarray}
E_\infty &=& \frac{1}{1 + \eta^2_3} \Big(\omega_2 + \omega_3 \, \eta^{2}_3 \Big)  
-  \frac{4}{(1 + \eta^2_3)^2} \frac{\mu^2_{23} \, \eta^{2}_3 }{\Omega_{23}} \, ,  
\end{eqnarray}
where we identify the energy surface of a two-level system with levels given by $\omega_2$ and $\omega_3$. The variable of this energy surface is $\eta_3$. The critical points are $\eta_3=0$, yielding an energy value $E=\omega_2$, and
\begin{equation}
\label{eq.eta.k3}
\eta^c_3  = \sqrt{\frac{4 \mu^2_{23} - (\omega_3-\omega_2)\Omega_{23}}{4 \mu^2_{23} + (\omega_3-\omega_2)\Omega_{23}}} \, ,
\end{equation}
with corresponding energy given by
\begin{equation}
E_{23} = \omega_2 -\frac{\left(4\,\mu_{23}^2- (\omega_3-\omega_2)\,\Omega_{23}\right)^2 }{16\, \Omega_{23}\,\mu_{23}^2} \, .
\end{equation}
The expressions coincide with the critical energy surface of a two-level system when the energy of the lower level ($\omega_2$) is different from zero. The latter critical point and its corresponding energy are valid in the parameter region where the inequality $4 \mu^2_{23} \geq (\omega_3-\omega_2)\Omega_{23}$ is satisfied. Again, the well known expression for the quantum phase transition from the normal to the superradiant regions for a 2 level system is obtained \cite{castanos11,castanos11b}. 

The phase diagram for the $3$-level system is established by considering first the equalities
\[
E_{\rm N}=E_{12} \,  , \quad E_{\rm N}=E_{13} \, , \quad  E_{\rm N}=E_{23} \, ,
\]
dividing the parameter space into regions, by
\begin{eqnarray*}
4 \, \mu^2_{12}  = \Omega_{12} \, \omega_2 \, , \quad 4 \, \mu^2_{13}  = \Omega_{13} \, \omega_3 \, ,  \\ 
 4 \, \mu^2_{23}= \Omega_{23} \Big( \sqrt \omega_3 + \sqrt \omega_2 \Big)^2 \, . 
\end{eqnarray*}
The first and second equations define bifurcation sets, where two different critical points coalesce to $(\varrho^c_2, \varrho^c_3)=(0,0)$, while the third equation defines a Maxwell set,
because the critical points are different but they have the same value of the energy.

We have further divisions of the parameter space, given by
\[
E_{12}=E_{13} \,  , \quad E_{12}=E_{23} \, , \quad  E_{13}=E_{23} \, ,
\]
The first equality gives
\begin{eqnarray*}
&& \mu^2_{12} = \frac{\Omega_{12}}{2 \, \Omega_{13} \, \mu^2_{13}} \Bigg( \mu^2_{13} - \frac{\Omega_{13} \, \omega_3}{4} \Bigg)^2  \quad \\
&\times& \Bigg( 1 + \frac{\mu^2_{13} \, \Omega_{13} \,  \omega_2 }{2 \, (\mu^2_{13} - \frac{\Omega_{13} \, \omega_3 }{4})^2} \pm \sqrt{1 + \frac{\mu^2_{13} \, \Omega_{13} \,  \omega_2}{(\mu^2_{13} - \frac{\Omega_{13} \, \omega_3 }{4})^2}} \ \Bigg) \, ,
\end{eqnarray*}
which must obey the constraints associated to the parameters $\mu_{12}$ and $\mu_{13}$ to define a critical point.  The second equality results in
\begin{eqnarray*}
&&\mu^2_{23} = \frac{\Omega_{23}}{2 \, \Omega_{12} \, \mu^2_{12}} \Bigg( \mu^2_{12} + \frac{\Omega_{12} \, \omega_2}{4} \Bigg)^2 \\
&\times& \Bigg( 1 + \frac{\mu^2_{12} \, \Omega_{12} \,  (\omega_3 - \omega_2)}{2 \, (\mu^2_{12} + \frac{\Omega_{12} \, \omega_2 }{4})^2} \pm \sqrt{1 + \frac{\mu^2_{12} \, \Omega_{12} \,  (\omega_3 - \omega_2)}{(\mu^2_{12} + \frac{\Omega_{12} \, \omega_2 }{4})^2}} \, \Bigg) \, ,
\end{eqnarray*}
and since we must have $\mu^2_{23} \geq (\omega_3 - \omega_2) \Omega_{23}/4$,   only the solution with the plus sign is in the collective regime. For the third equality one has the first result given above by making the replacements $\mu_{12} \to \mu_{13}$, $ \Omega_{12} \to  \Omega_{13}$, $\omega_2 \to \omega_3$, and $\omega_3 \to \omega_2$.

The curves above determine the region boundaries where a quantum phase transition between two superradiant regions takes place. They correspond to transitions between the pairs of photonic mode: $(\Omega_{12}, \Omega_{13})$,  $(\Omega_{12}, \Omega_{23})$ and $(\Omega_{13}, \Omega_{23})$ respectively. Due to the nature of the critical points, they are all Maxwell sets.  

The expectation values for the matter and field observables with respect to the variational states may be obtained at the critical points. In table~\ref{t3niv} we show them for the different regions in parameter space. We have labelled by $S_{ij}$ the superradian region where the mode $\Omega_{ij}$ dominates. In all cases these quantities are proportional to the number of particles, and we choose to show them per particle.

\begin{table*}
\begin{center}
\caption{Expectation values and dispersions for the number of photons and atomic populations per particle for the variational ground state in $3$-level systems, in the different regions of the phase diagram. We have defined $\phi_{jk}\equiv\phi_{k}-\phi_{j}$, and the condition for a minimum is $\phi_{jk}=0,\,\pi$. In the text we have taken $\omega_{1}=0$ without loss of  generality.}
\label{t3niv}
\vskip2mm
\begin{tabular}{l |c |c|c|c}
 & \ {N} \ &  {$S_{12}$} &  {$S_{23}$} &  {$S_{13}$} \\ \hline\hline &&&\\[-2mm] 
$\langle \bm{\nu}_{12}\rangle$ & 0 & $\displaystyle\frac{\mu_{12}^2}{\Omega_{12}^2}\left(1 - \frac{(\omega_2-\omega_{1})^2\,\Omega_{12}^2}{16\,\mu_{12}^4}\right)$ & 0 & 0 \\[4mm]
$\langle \bm{\nu}_{23}\rangle$ & 0 & 0 & $\displaystyle\frac{\mu_{23}^2}{\Omega_{23}^2}\left(1 - \frac{(\omega_3-\omega_2)^2\,\Omega_{23}^2}{16\,\mu_{23}^4}\right)$  &  0\\[4mm]
$\langle \bm{\nu}_{13}\rangle$ & 0 & 0 & 0  & $\displaystyle\frac{\mu_{13}^2}{\Omega_{13}^2}\left(1 - \frac{(\omega_3-\omega_{1})^2\,\Omega_{13}^2}{16\,\mu_{13}^4}\right)$ \\[4mm]
\hline&&&& \\[1mm]
$\langle \bm{A}_{11}\rangle$ & $1$ & $\displaystyle\frac{1}{2}\left(1 + \frac{(\omega_2-\omega_{1})\,\Omega_{12}}{4\,\mu_{12}^2}\right)$ & 0 & $\displaystyle\frac{1}{2} \left(1+ \frac{(\omega_3-\omega_{1})\,\Omega_{13}}{4\,\mu_{13}^2}\right)$\\[4mm]
$\langle \bm{A}_{22}\rangle$ & 0 & $\displaystyle\frac{1}{2}\left(1 - \frac{(\omega_2-\omega_{1})\,\Omega_{12}}{4\,\mu_{12}^2}\right)$ &  $\displaystyle\frac{1}{2}\left(1 + \frac{(\omega_3-\omega_2)\,\Omega_{23}}{4\,\mu_{23}^2}\right)$ & 0 \\[4mm]
$\langle \bm{A}_{33}\rangle$ & 0 & 0 & $\displaystyle\frac{1}{2}\left(1 - \frac{(\omega_3-\omega_2)\,\Omega_{23}}{4\,\mu_{23}^2}\right)$ & $\displaystyle\frac{1}{2}\left(1 - \frac{(\omega_3-\omega_{1})\,\Omega_{13}}{4\,\mu_{13}^2}\right)$ \\[4mm]
$\langle \bm{A}_{12}\rangle$ & 0 &$\displaystyle\frac{1}{2}\left(1-\frac{(\omega_{2}-\omega_{1})^{2}\,\Omega_{12}^{2}}{16\,\mu_{12}^{4}}\right)^{1/2}\,e^{i\,\phi_{12}}$ & 0 & 0 \\[4mm]
$\langle \bm{A}_{13}\rangle$ & 0 & 0 & 0 & $\displaystyle\frac{1}{2}\left(1-\frac{(\omega_{3}-\omega_{1})^{2}\,\Omega_{13}^{2}}{16\,\mu_{13}^{4}}\right)^{1/2}\,e^{i\,\phi_{13}}$  \\[4mm]
$\langle \bm{A}_{23}\rangle$ & 0 & 0 & $\displaystyle\frac{1}{2}\left(1-\frac{(\omega_{3}-\omega_{2})^{2}\,\Omega_{23}^{2}}{16\,\mu_{23}^{4}}\right)^{1/2}\,e^{i\,\phi_{23}}$ & 0 \\[4mm]
\hline&&&& \\[1mm]
$(\Delta \bm{A}_{11})^2$ &0&$\displaystyle\frac{1}{4} \left(1- \frac{(\omega_2-\omega_{1})^2\,\Omega_{12}^2}{16\,\mu_{12}^4}\right)$& 0 & $\displaystyle\frac{1}{4}\left(1 - \frac{(\omega_3-\omega_{1})^2\,\Omega_{13}^2}{16\,\mu_{13}^4}\right)$ \\[4mm]
$(\Delta\bm{A}_{22})^2$ &0&$\displaystyle\frac{1}{4} \left(1- \frac{(\omega_2-\omega_{1})^2\,\Omega_{12}^2}{16\,\mu_{12}^4}\right)$& $\displaystyle\frac{1}{4}\left(1 - \frac{(\omega_3-\omega_2)^2\Omega_{23}^2}{16\,\mu_{23}^4}\right)$ & 0 \\[4mm]
$(\Delta\bm{A}_{33})^2$ &0&0& $\displaystyle\frac{1}{4}\left(1 - \frac{(\omega_3-\omega_2)^2\Omega_{23}^2}{16\,\mu_{23}^4}\right)$ & $\displaystyle\frac{1}{4}\left(1 - \frac{(\omega_3-\omega_{1})^2\Omega_{13}^2}{16\,\mu_{13}^4}\right)$\\[4mm]
\end{tabular}
\end{center}
\end{table*}

By means of the critical points one can give explicitly the corresponding variational ground states for the different regions in parameter space: 
\begin{eqnarray}
|\Psi_{\rm N}\rangle &=& \frac{1}{\sqrt{N_a!}}{\bm{b}_1^\dag}^{N_a}\,|0\rangle_{M} \,\otimes\, |0\rangle_{F}\,,\\[3mm]
|\Psi_{1k}\rangle &=& \frac{1}{\sqrt{N_a!}}\left(\frac{\bm{b}_1^\dag + \varrho_k^c \bm{b}_k^\dag}{\sqrt{1+{\varrho_k^c}^2}}\right)^{N_a}\,|0\rangle_{M} \nonumber\\[3mm]
&\otimes&\exp(-N_a\,r_{1k}^{c\, 2}/2)\,\exp(\sqrt{N_a}\,r_{1k}^{c}\,\bm{a}_{1k}^{\dagger})
\,|0\rangle_{F}\,,\\[3mm]
|\Psi_{23}\rangle &=&  \frac{1}{\sqrt{N_a!}}\left(\frac{\bm{b}_2^\dag + \eta_3^c \bm{b}_3^\dag}{\sqrt{1+{\eta_3^c}^2}}\right)^{N_a}\,|0\rangle_{M}\nonumber\\[3mm]
&\otimes&\exp(-N_a\,\tilde{r}_{23}^{c\, 2}/2)\,\exp(\sqrt{N_a}\,\tilde{r}_{23}^{c}\,\bm{a}_{23}^{\dagger})
\,|0\rangle_{F}\,,
\end{eqnarray}
%
where in the second equation $k=2,\,3$, and the expressions for $r_{23}^c$, $\varrho_2^c$, $\varrho_3^c$ and $\eta_3^c$ are given by equations (\ref{eq.rho.c}), (\ref{eq.varrho.k2}), (\ref{eq.varrho.k3}) and (\ref{eq.eta.k3}), respectively. The value $\tilde r_{23}^c$ is obtained in the limit $\varrho_{2}^c\to\infty$ in the expression for $r_{23}^c$ assuming $\varrho_{3}^c=\eta_3 \,\varrho_{2}^c$, and yields 
\begin{eqnarray}
\tilde r_{23}^c = \lim_{\varrho_{2}^c\to\infty}\,r_{23}^c= \frac{2\,\mu_{23}\,\eta_{3}^c}{(1+{\eta_3^c}^2)\Omega_{23}}\,.
\end{eqnarray}

\begin{figure}
\begin{center}
\includegraphics[width=\linewidth]{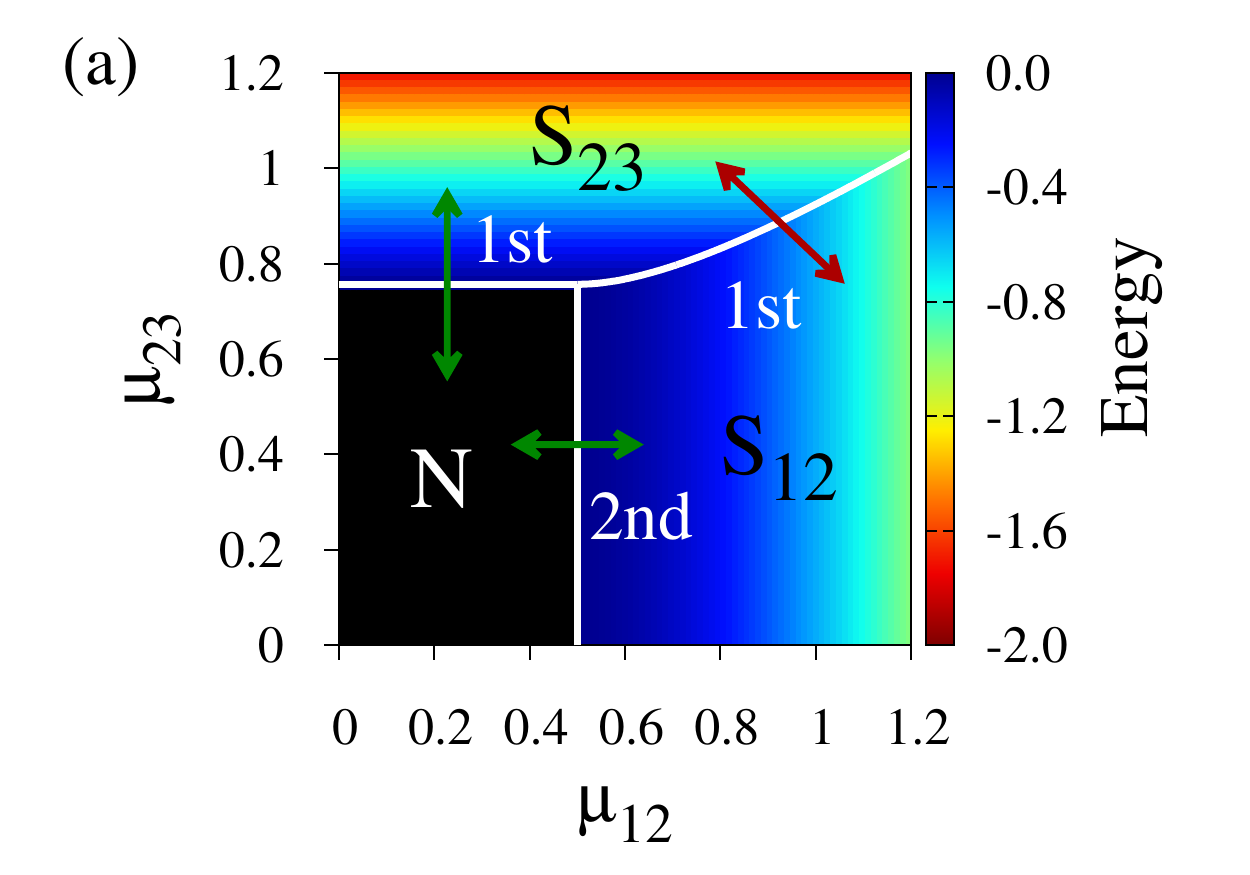}
\includegraphics[width=\linewidth]{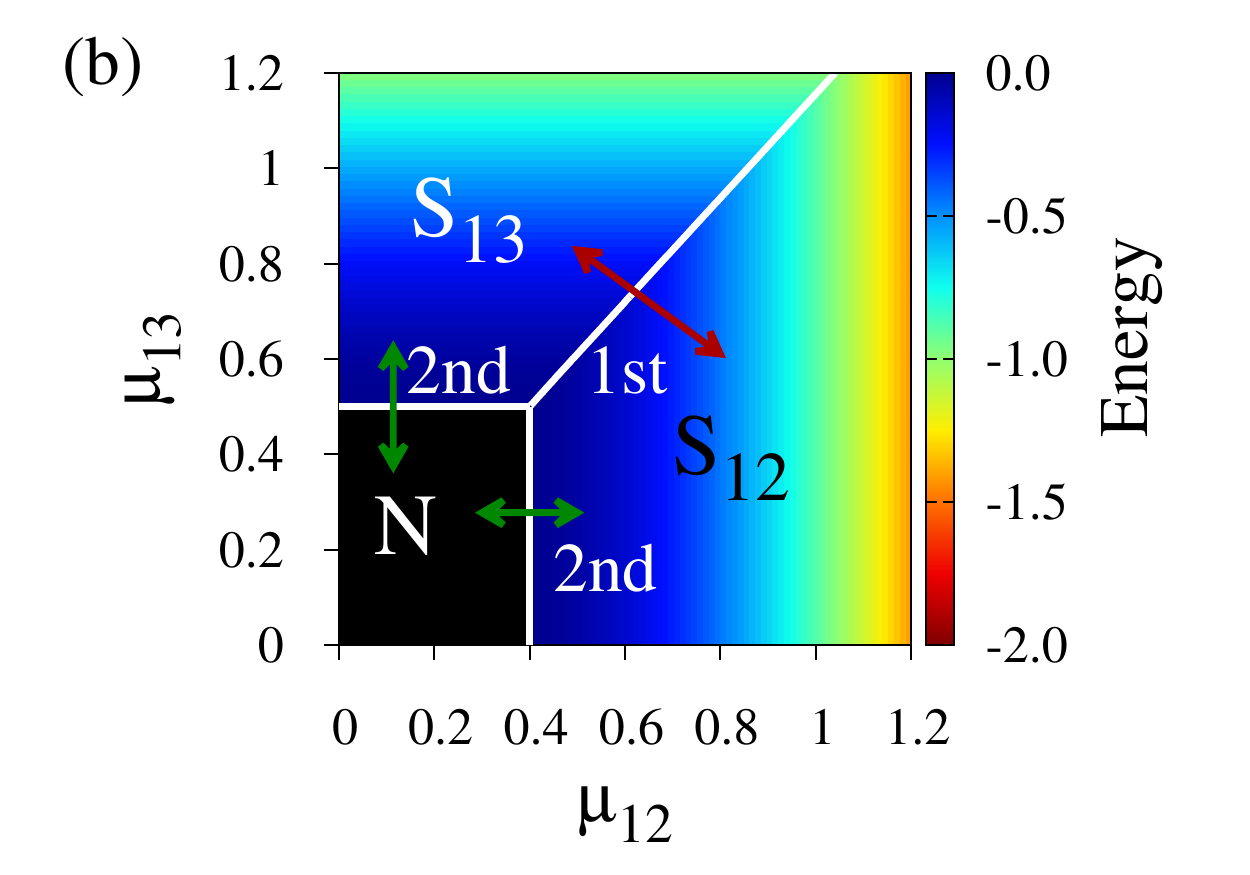}
\includegraphics[width=\linewidth]{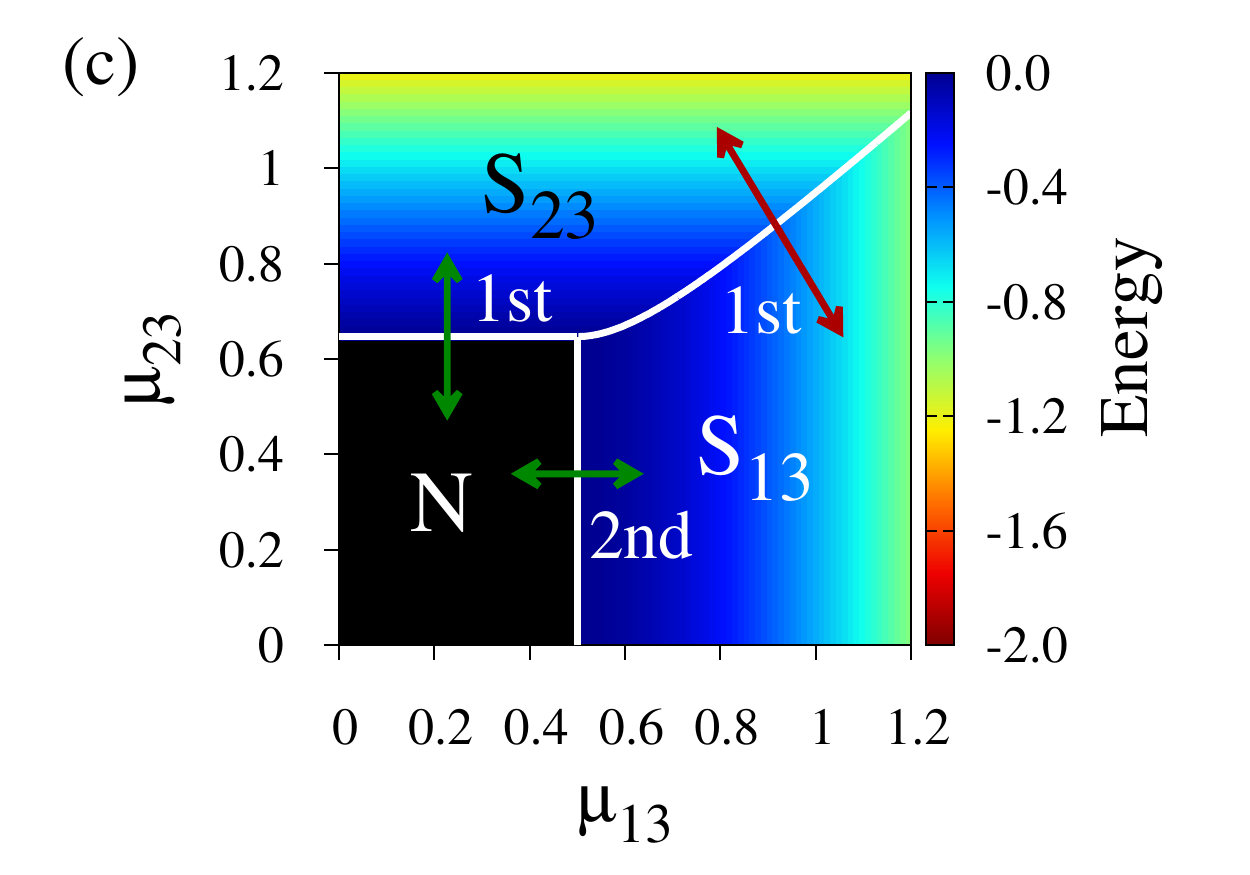}
\caption{(color online) Energies and phase diagrams for atoms in the $\Xi$-, $V$-, and $\Lambda$-configurations. The separatrices (white lines) and the order of the transitions are shown. The normal regions are labelled by $N$  (black). The collective regime is divided into the regions:
(a) $S_{12}$ and $S_{23}$ corresponding to the active modes $\Omega_{12}$ and $\Omega_{23}$ respectively, shown here for $\Omega_{12} = 1,\, \Omega_{23} = 0.5,\, \omega_2 = 1$ and $\omega_3 = 1.3$;
(b) $S_{12}$ and $S_{13}$ corresponding to the active modes $\Omega_{12}$ and $\Omega_{13}$ respectively, with $\Omega_{13} = 1,\, \Omega_{12} = 0.8,\, \omega_2 = 0.8$ and $\omega_3 = 1$; and
(c) $S_{13}$ and $S_{23}$ corresponding to the active modes $\Omega_{13}$ and $\Omega_{23}$ respectively, with $\Omega_{13} = 1,\, \Omega_{23} = 0.8,\, \omega_2 = 0.2$ and $\omega_3 = 1$.}\label{enXVL}
\end{center}
\end{figure}

From these general results for $3$-level systems we can easily obtain the phase diagrams for the different atomic configurations. For the $\Xi$-atomic configuration ($\mu_{13}=0$) one has the critical energy surfaces $E_N,\, E_{12},\, E_{23}$, from which the phase diagram can be obtained and is displayed in figure~\ref{enXVL}\,(a).  

For the $V$-configuration ($\mu_{23}=0$) we have critical energy surfaces $E_N,\, E_{12},\, E_{13}$, and the corresponding phase diagram is plotted in figure~\ref{enXVL}\,(b).

Finally, for the $\Lambda$-configuration ($\mu_{12}=0$) the minimum energy surfaces are $E_N,\, E_{13},\, E_{23}$, and the quantum phase diagram is shown in figure~\ref{enXVL}\,(c).

In general we may write
\begin{eqnarray}
E_{\Xi min} &=& \min\{E_{{\rm N}},\, E_{12},\,E_{23} \} \, , \nonumber \\
E_{V min} &=& \min\{E_{{\rm N}},\, E_{12},\,E_{13} \} \, , \nonumber \\
E_{\Lambda min} &=& \min\{E_{{\rm N}},\, E_{13},\,E_{23} \}  \, ,
\end{eqnarray}
where $E_N$ exists for all values of $\mu_{ij}$, $E_{12}$ is independent of $\mu_{13}$ and $\mu_{23}$, and so on. Other critical values for $\varrho_{2}$ and $\varrho_{3}$ exist; however, in all cases they provide higher energy values. Notice that since the solutions for $E_{12}$, $E_{13}$, and $E_{23}$ only have emission or absorption of photons associated to the modes $\Omega_{12}$, $\Omega_{13}$ and $\Omega_{23}$, respectively,  the polychromatic phase diagram for the variational ground state is naturally divided into monochromatic subregions, labelled by $S_{ij}$ in Fig.~\ref{enXVL}.

In figure \ref{enXVL}\,(a), the minimum energy surface as a function of the dipolar intensities $\mu_{12},\,\mu_{23}$ is presented. The normal region $E_{N}=0$ is shown in black, while the collective regions, with values $E_{\Xi min}<0$, appear in a color graded scale. The separatrix (white lines) is defined by the set of points $(\mu_{12},\mu_{23})$ where $E_{\rm N} = E_{12}$ and $E_{\rm N} = E_{23}$, indicating the borders between the normal and collective regions, together with the points where $E_{12} = E_{23}$ dividing the collective regimen into monochromatic regions, corresponding to the two electromagnetic modes.

According to the Ehrenfest classification~\cite{gilmore93}, a phase transition is of order $j$, if $j$ is the lowest non-negative integer for which 
\begin{equation*}
\lim_{\epsilon \to 0} \frac{\partial^j E^c}{\partial \mu^j} \Bigg |_{\mu=\mu_0+\epsilon} \neq \lim_{\epsilon \to 0} \frac{\partial^j E^c}{\partial \mu^j} \Bigg |_{\mu=\mu_0-\epsilon} \, ,
\end{equation*}
where $\mu$ represents a dipolar intensity parameter. In our case the ground state undergoes a second order phase transitions when  the dipolar intensities go from region~{\rm N} to region~{$S_{12}$} and vice versa, but first order phase transitions occur when crossing from region~{\rm N} to region~{$S_{23}$} and from region~{$S_{12}$} to region~{$S_{23}$}.     
%
\begin{figure}[h!]
\begin{center}
\includegraphics[width=0.8\linewidth]{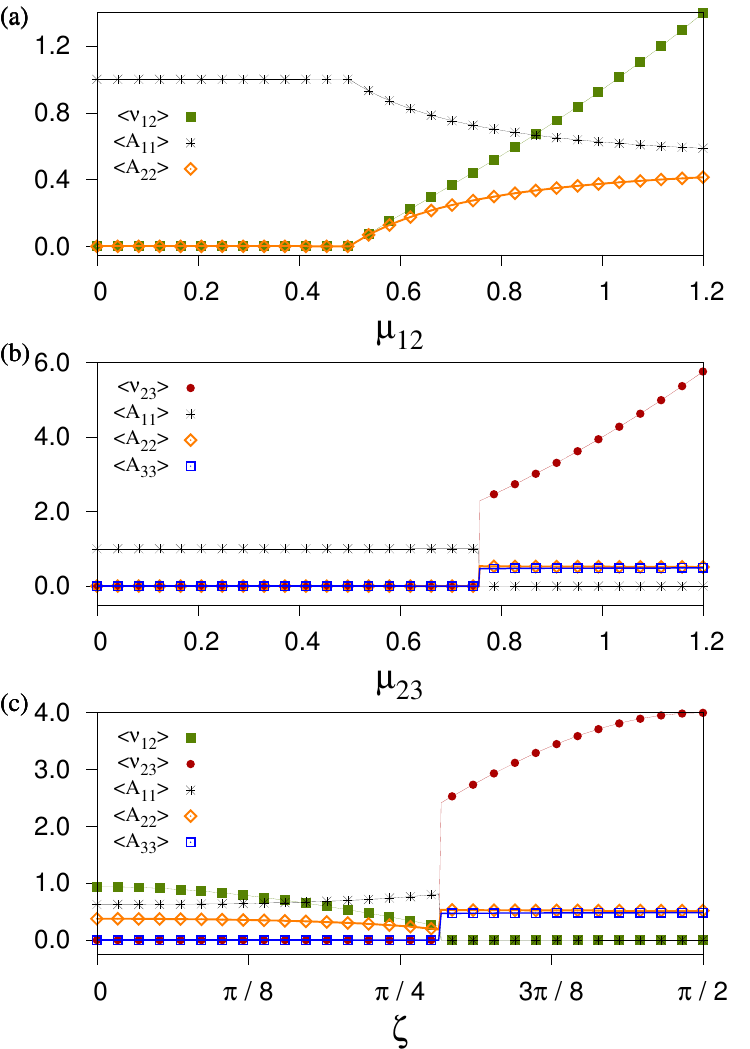}
\caption{(color online)  behavior of the expectation values of the photon and atomic level populations, as functions of the interaction parameters. See text for details.}\label{expectX}
\end{center}
\end{figure}

As an example of the behavior of the expectation values given in Table~\ref{t3niv}, 
figure~\ref{expectX}(a)  shows  $\langle\bm{\nu}_{12}\rangle = \langle\bm{a}_{12}^\dag\bm{a}_{12}\rangle$ and $\langle\bm{A}_{kk}\rangle$ ($k=1, \, 2$) as functions of $\mu_{12}$ for  $\mu_{23}=0$; figure~\ref{expectX}(b) shows $\langle\bm{\nu}_{23}\rangle = \langle\bm{a}_{23}^\dag\bm{a}_{23}\rangle$ and $\langle\bm{A}_{kk}\rangle$ ($k=1, \, 2, \, 3$) as functions of $\mu_{23}$ for  $\mu_{12}=0$; and figure~\ref{expectX}(c) shows $\langle\bm{\nu}_{jk}\rangle = \langle\bm{a}_{jk}^\dag\bm{a}_{jk}\rangle$ and $\langle\bm{A}_{kk}\rangle$ ($j<k=1, \, 2$) as functions of $\zeta$, where we have parametrised the dipole interactions in the form $\mu_{12}= \mu \cos\zeta$,  $\mu_{23}= \mu \sin\zeta$, $\mu=1$, and $0\leq\zeta\leq\pi/2$, for a trajectory going through regions {$S_{12}$} and {$S_{23}$}. The breakpoints and discontinuities inherited from $E_{\Xi min}$ and its derivatives at the loci of a phase transition are evident. For first order transitions the expectation values exhibit a discontinuity. We have taken $\Omega_{12}=1$, $\Omega_{23}=0.5$, $\omega_2=1$, $\omega_3=1.3$. The same results are obtained for all values of $\mu_{12}$ that satisfy $4\,\mu_{12}^2 - \omega_2\,\Omega_1<0$, or values of $\mu_{23}$ satisfying $4\,\mu_{23}^2 -(\omega_3-\omega_2)\,\Omega_{23}<0$ (see Fig.~\ref{enXVL}\,(a)).  

Figures~\ref{enXVL}\,(b) and ~\ref{enXVL}\,(c) show maps of the minimum energy surface for atoms in the $V$- and $\Lambda$-configurations. The normal region is shown in black, and the collective region in a color graded scale.
For the $V$ case we have second order transitions when the system goes from  {\rm N} to {$S_{12}$} and from  {\rm N} to {$S_{13}$}, and first order transitions  when it goes from  {$S_{12}$} to {$S_{13}$}.  A similar analysis of the physical quantities (number of photons and atomic populations) to that of figure~\ref{expectX} will show that they all vary continuously as functions of $\mu_{12}$ with $4\, \mu_{13}^2-\omega_3\,\Omega_{13}<0$ and as functions of $\mu_{13}$ with  $4\,\mu_{12}^2 - \omega_2\,\Omega_{12}<0$, because of the second order of the transition,  while they exhibit discontinuities when the system goes from region {$S_{12}$}  to {$S_{13}$} as this is a first order transition.
For the $\Lambda$ case, we have $\omega_2>\omega_1$, the phase diagram for atoms in this $\Lambda$-configuration resembles the case of the $\Xi$-configuration (compare figures~\ref{enXVL}\,(a) and \ref{enXVL}\,(c)). The expectation values of physical quantities vary continuously when the system goes from {\rm N} to $S_{13}$, where a second order transition occurs, and they present discontinuities when the systems goes from {\rm N} to $S_{23}$ and from $S_{13}$ to $S_{23}$, where the system undergoes a first order phase transition.

All the results for $3$-level atoms show that the polychromatic minimum energy surface divides itself into monochromatic regions where only one mode of electromagnetic field contributes to the variational ground state.  Also, as first order transitions are related to discontinuities in at least one physical quantity, these critical points along the separatrix form a Maxwell set, {\em i.e.}, the corresponding critical points in the transition  $\varrho^{c} = 0$ and $\varrho^{c}\to\infty$ provide the same minimum energy value, whereas the critical points where second order transitions occur correspond to bifurcations.

\section{$4$-level atoms}\label{s.4l}

Taking $n=4$ in the expression of the variational energy surface~(\ref{E-critica}) one has
\begin{eqnarray}
E &=& \frac{1}{1 + R_0^2} \Big(\omega_2 \, \varrho^{ c \, 2}_2 
+ \omega_3 \, \varrho^{ c \, 2}_3 + \omega_4 \varrho^{ c \, 2}_4 \Big)  \nonumber \\ 
&-&  \frac{4}{(1 + R_0^2)^2} \frac{\mu^2_{23}  \, \varrho^{ c \, 2}_2\, \varrho^{ c \, 2}_3 }{\Omega_{23}}  
 - \frac{4}{(1 + R_0^2)^2} \frac{\mu^2_{12} \, \varrho^{ c \, 2}_2}{\Omega_{12}} \nonumber \\ 
 &-& \frac{4}{(1 + R_0^2)^2} \frac{\mu^2_{13} \, \varrho^{ c \, 2}_3}{\Omega_{13}} 
 - \frac{4}{(1 + R_0^2)^2} \frac{\mu^2_{14} \, \varrho^{ c \, 2}_4}{\Omega_{14}} \nonumber \\
 &-&  \frac{4}{(1 + R_0^2)^2} \frac{\mu^2_{24}  \, \varrho^{ c \, 2}_2\, \varrho^{ c \, 2}_4 }{\Omega_{24}} \nonumber \\
 &-&  \frac{4}{(1 + R_0^2)^2} \frac{\mu^2_{34}  \, \varrho^{ c \, 2}_3\, \varrho^{ c \, 2}_4 }{\Omega_{34}}  \, .  
\end{eqnarray}
Again, $(\varrho^{ c}_2,\varrho^c_3, \varrho^c_4) =(0,0,0)$ is a critical point which yields an energy surface ${E}^c_{\rm N}=0$.

By considering $\varrho^c_4 =0$ and $\varrho^c_k \neq 0$ with $k=2,3$, the energy surface takes the form of a $3$-level system
\begin{eqnarray}\label{eq.37}
E &=& \frac{1}{1 + R_0^2} \Big(\omega_2 \, \varrho^{ c \, 2}_2 + \omega_3 \, \varrho^{ c \, 2}_3  \Big) -  \frac{4}{(1 + R_0^2)^2} \frac{\mu^2_{23}  \, \varrho^{ c \, 2}_2\, \varrho^{ c \, 2}_3 }{\Omega_{23}}  \nonumber \\
 &-& \frac{4}{(1 + R_0^2)^2} \frac{\mu^2_{12} \, \varrho^{ c \, 2}_2}{\Omega_{12}} - \frac{4}{(1 + R_0^2)^2} \frac{\mu^2_{13} \, \varrho^{ c \, 2}_3}{\Omega_{13}}   \, . 
\end{eqnarray}
By further taking $\varrho^c_4 =0$ and $\varrho^c_3 =0$ we have
\begin{equation}
\varrho^{c}_2 = \sqrt{\frac{4 \mu^2_{12} - \Omega_{12} \, \omega_2}{4 \mu^2_{12} + \Omega_{12} \, \omega_2}} \, , \quad   \, {E}_{12} = -\frac{(- 4 \, \mu^2_{12} +  \Omega_{12} \, \omega_2)^2}{16 \mu^2_{12} \, \Omega_{12}} \, ,
\end{equation}
while for $\varrho^c_4 =0$ and $\varrho^c_2 =0$ one gets
\begin{equation}
\varrho^{c}_3 = \sqrt{\frac{4 \mu^2_{13} - \Omega_{13} \, \omega_3}{4 \mu^2_{13} + \Omega_{13} \, \omega_3}} \, , \quad   \, {E}_{13} = -\frac{(- 4 \, \mu^2_{13} +  \Omega_{13} \, \omega_3)^2}{16 \mu^2_{13} \, \Omega_{13}} \, . 
\end{equation}
where the previous expressions exist only for values of the dipolar strength $ 4 \, \mu^2_{1k} \geq  \Omega_{1k} \, \omega_k$, with $k=2,3$. 

In the limit when the variables go to infinity, one has $\varrho^c_4 =0$, $\varrho^c_2 \to \infty$,  
$\varrho^c_3=\eta_3 \, \varrho^c_2$, and the critical point given by
\begin{eqnarray}
\label{4lev}
\eta^c_3&=& \sqrt{\frac{4 \mu^2_{23} - (\omega_3-\omega_2)\Omega_{23}}{4 \mu^2_{23} + (\omega_3-\omega_2)\Omega_{23}}} \, , 
\end{eqnarray}
from which the energy surface takes the expression
\begin{eqnarray}
{E}_{23} = \omega_2 -\frac{\left(4\,\mu_{23}^2- (\omega_3-\omega_2)\,\Omega_{23}\right)^2 }{16\, \Omega_{23}\,\mu_{23}^2} \, , \quad 
\end{eqnarray}
which is valid in the parameter region where the inequality $4 \mu^2_{23} \geq (\omega_3-\omega_2)\Omega_{23}$ is satisfied.  
The expression coincides with the critical energy surface of a two-level system. One again, the well known expression for the quantum phase transition from the normal to the superradiant regions for a 2-level system is obtained.

\begin{figure}
\begin{center}
\includegraphics[width=0.9\linewidth]{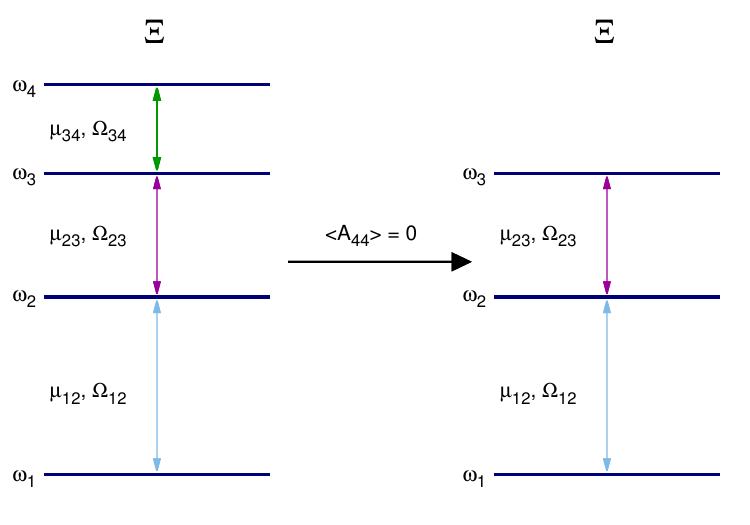}
\caption{(color online) Under the condition $\langle \bm{A}_{44}\rangle = 0$ for the ground state, a system of $4$-level atoms in the $\Xi$-configuration is reduced to $3$-level atoms in the $\Xi$-configuration. }\label{fX4toX3}
\end{center}
\end{figure}

\begin{figure}
\begin{center}
\includegraphics[width=0.9\linewidth]{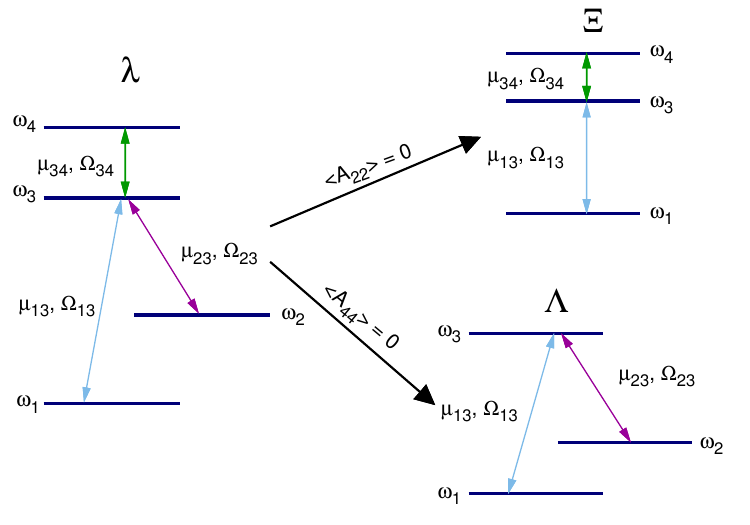}\\[5mm]
\includegraphics[width=0.9\linewidth]{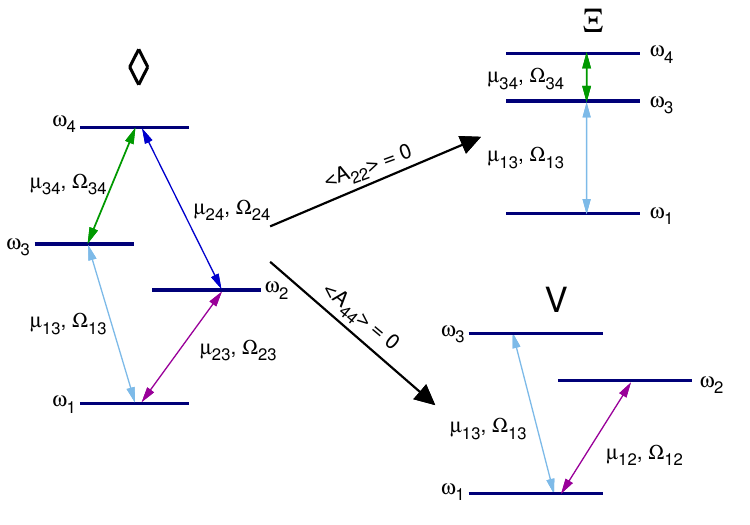}
\caption{(color online) Schematic procedure to obtain, by a simple consideration ($\langle\bm{A}_{kk}\rangle=0$), a $3$-level systems from a $4$-level system. The $\lambda$-configuration is reduced to the $\Xi$- or $\Lambda$-configuration and  the $\diamondsuit$-configuration is reduced to the $\Xi$- or $V$-configuration. In both cases the simple condition on the ground state is given by $\langle\bm{A}_{22}\rangle = 0$ and $\langle\bm{A}_{44}\rangle = 0$, respectively.}\label{f4to3}
\end{center}
\end{figure}

$4$-level systems have many configurations named by the atomic physics community by $\Xi,\,\lambda,\,{\rm y},\,N,\,\diamondsuit,\,\dots$, and the number of modes yielding transitions between pairs of atomic levels depend on the considered configuration. 
For $4$-level atoms in the $\Xi$-configuration, to take $\varrho^c_4=0$ is equivalent to decouple the fourth level from the system and the population of this level therefore vanishes.  Thus the system is reduced to $3$-level atoms in the same configuration as shown schematically in figure~\ref{fX4toX3}. The $\lambda$- and $\diamondsuit$-configurations, under the same condition,  are reduced to the $3$-level atomic configurations $\Lambda$ and $V$ respectively, as shown in figure~\ref{f4to3}. On the other hand, under the condition $\varrho^c_2=0$ (to be considered below) the $\lambda$- and $\diamondsuit$-configurations are reduced to $3$-level atomic $\Xi$-configurations as shown in the same figure. 
%
%

By considering the case $\varrho^c_3 =0$ and $\varrho^c_k \neq 0$ with $k=2,\,4$, the corresponding energy surface takes the form of a $3$-level system by making the replacement of the label $3 \to 4$ in equation~(\ref{eq.37})
\begin{eqnarray}
&&\omega_3 \to \omega_4 \, , \ \Omega_{13} \to \Omega_{14} \, , \ \Omega_{23} \to \Omega_{24} \, , \nonumber \\
&&\varrho^c_3 \to \varrho^c_4 \, , \ \mu_{13} \to \mu_{14} \, , \ \mu_{23} \to \mu_{24} \, . 
\end{eqnarray}
For the case $\varrho^c_3 =0$ and $\varrho^c_4 =0$ one finds the result already considered above. Then if we take $\varrho^c_3 =0$ and $\varrho^c_2 =0$, the critical points and energy surface are given by
\begin{eqnarray}
\label{4lev2}
\varrho^{c}_4 &=& \sqrt{\frac{4 \mu^2_{14} - \Omega_{14} \, \omega_4}{4 \mu^2_{14} + \Omega_{14} \, \omega_4}} \, , 
\end{eqnarray}
and
\begin{eqnarray}
  \, {E}_{14} = -\frac{(- 4 \, \mu^2_{14} +  \Omega_{14} \, \omega_4)^2}{16 \mu^2_{14} \, \Omega_{14}} \, ,
\end{eqnarray}
for values satisfying $4\,\mu_{14}^2-\Omega_{14}\,\omega_4\geq 0$.

In the limit  $\varrho^c_2 \to \infty$ and $\varrho^c_4=\eta_4 \, \varrho^c_2$, one gets 
\begin{eqnarray}
\eta^c_4&=& \sqrt{\frac{4 \mu^2_{24} - (\omega_4-\omega_2)\Omega_{24}}{4 \mu^2_{24} + (\omega_4-\omega_2)\Omega_{24}}} \, , 
\end{eqnarray}
and the energy surface is
\begin{eqnarray}
{E}_{24} = \omega_2 -\frac{\left(4\,\mu_{24}^2- (\omega_4-\omega_2)\,\Omega_{24}\right)^2 }{16\, \Omega_{24}\,\mu_{24}^2} \, ,
\end{eqnarray}
with $4\,\mu_{24}^2-(\omega_4-\omega_2)\Omega_{24}\geq 0$.

Finally, for $\varrho^c_2 =0$ and $\varrho^c_k \neq 0$ with $k=3,4$, the corresponding energy surface takes the form of a $3$-level system with the replacements of the labels $2 \to 3$ and $3 \to 4$.  Then if we consider $\varrho^c_4 = 0$, or $\varrho^c_3 = 0$, we have the results already discussed above, i.e., ${E}_{13}$ and ${E}_{14}$.

In the limit when the variables go to infinity, one has $\varrho^c_2 =0$, $\varrho^c_3 \to \infty$,  
$\varrho^c_4=\bar{\eta}_4 \, \varrho^c_3$, and thus the critical points are given by
\begin{eqnarray}
\label{4lev3}
\bar{\eta}^c_4&=& \sqrt{\frac{4 \mu^2_{34} - (\omega_4-\omega_3)\Omega_{34}}{4 \mu^2_{34} + (\omega_4-\omega_3)\Omega_{34}}} \, , 
\end{eqnarray}
with 
\begin{eqnarray}
{E}_{34} = \omega_3 -\frac{\left(4\,\mu_{34}^2- (\omega_4-\omega_3)\,\Omega_{34}\right)^2 }{16\, \Omega_{34}\,\mu_{34}^2} \, , 
\end{eqnarray}
when $4\,\mu_{34}^2-(\omega_4-\omega_3)\Omega_{34}\geq 0$ is fulfilled.

The phase diagram for a $4$-level system is established by considering the equality of the different energy surfaces, that is, from
\begin{eqnarray}
{E}^c_{{\rm N}} &=& {E}^c_{12} \,  , \quad {E}^c_{\rm N}={E}^c_{13} \, , \quad  {E}^c_{\rm N}={E}^c_{23} \, ,  \nonumber \\
{E}^c_{12} &=& {E}^c_{13} \,  , \quad {E}^c_{12}={E}^c_{23} \, , \quad  {E}^c_{13}={E}^c_{23} \, ,
\end{eqnarray}
giving the expressions obtained already in the $3$-level case. Therefore we will have a separatrix formed with bifurcation and Maxwell sets. 

The remaining conditions are associated to the new critical energy surfaces by establishing the equalities 
\begin{eqnarray}
{E}^c_{{\rm N}} &=& {E}^c_{14} \,  , \quad {E}^c_{\rm N}={E}^c_{24} \, , \quad  {E}^c_{\rm N}={E}^c_{34} \,, \nonumber \\
{E}^c_{{\rm 12}} &=& {E}^c_{14} \,  , \quad {E}^c_{12}={E}^c_{24} \, , \quad  {E}^c_{12}={E}^c_{34} \,,  \nonumber \\
{E}^c_{{\rm 13}} &=& {E}^c_{14} \,  , \quad {E}^c_{13}={E}^c_{24} \, , \quad  {E}^c_{13}={E}^c_{34} \,,  \nonumber \\
{E}^c_{{\rm 23}} &=& {E}^c_{14} \,  , \quad {E}^c_{23}={E}^c_{24} \, , \quad  {E}^c_{23}={E}^c_{34} \,, \nonumber \\
 {E}^c_{14} &=& {E}^c_{24} \, , \quad  {E}^c_{14}={E}^c_{34} \,  , \quad
 {E}^c_{{\rm 24}} ={E}^c_{34} \,. \nonumber 
\end{eqnarray}
The previous results determine the total number of borders associated to $4$-level atoms, however some of them are not allowed by the selection rules of the dipolar interaction.  To see this, we consider the 
$\Xi$-configuration where one has $\mu_{13}=\mu_{14}=\mu_{24}=0$.  Then the quantum phase diagram is divided into $6$ regions separating the normal and collective regimes
\begin{eqnarray}
{E}^c_{{\rm N}} &=& {E}^c_{12} \,  , \quad {E}^c_{\rm N}={E}^c_{23}  \, ,  \quad
 {E}^c_{{\rm N}} = {E}^c_{34} \,  , \nonumber \\
 {E}^c_{12} &=& {E}^c_{23} \,  ,\quad
{E}^c_{{\rm 12}} = {E}^c_{34} \,  , \quad {E}^c_{23}={E}^c_{34} \, ,
\end{eqnarray}
whose explicit expressions and existence condition were established previously. 

\begin{figure}
\begin{center}
\includegraphics[width=\linewidth]{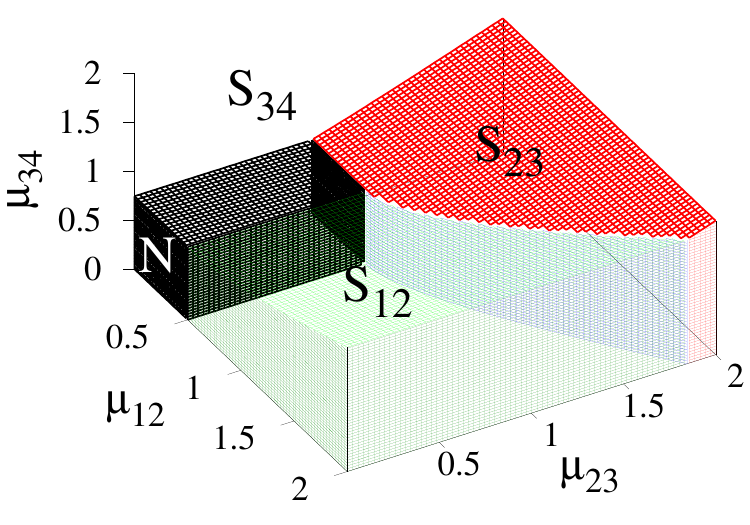}
\caption{(color online) Phase diagram as a function of the dipolar intensities for atoms of $4$-levels in the $\Xi$-configuration.  Parameters are: $\Omega_{12} = 1,\,  \Omega_{23} = 0.7,\, \Omega_{34}=0.3,\,  \omega_1 = 0,\, \omega_2 = 1,\,\omega_3=1.7$ and $\omega_4 = 2$.}\label{sepX4}
\end{center}
\end{figure}

The minimum energy surface for the this configuration of $4$ levels is given by 
\begin{equation}
E_{\Xi min} = \min\{E_{{\rm N}},\, E_{12},\, E_{23} \, , E_{34} \}  \, ,
\end{equation}
where, again, one must take into account the regions in parameter space where they are valid. In figure~\ref{sepX4} the separatrix of this system is shown. Clearly the energy surface is divided in a normal region and three collective regions, where in each of the latter only a monochromatic electromagnetic field contributes to the ground state. One finds second order transitions when the system goes from region~{N} (normal region) to region~${S_{12}}$, while for other crossings the transitions are of first order.

%

\section{$n$-level atoms}\label{s.gen}

We may generalise the above results to $n$-level atoms in dipolar interaction with $\ell$ modes of electromagnetic field.

We have found three types of minimum energy surfaces:
\begin{itemize}
\item $E_{N}=0$, corresponding to critical points $\varrho^c_{s}=0$ for $s=2,\,3,\dots,\,n$. 
\item $E_{1k}$, when $\Omega_{1k}$ promotes the transition $\omega_1 \rightleftharpoons \omega_k$ with $\varrho^c_{s}=0$ for all $s\neq k$,  given by
\begin{equation}\label{eq.E1kp}
E_{1k} = -\frac{\left(\omega_{k}\,\Omega_{1k }-4\,\mu_{1k}^2\right)^2}{16\,\Omega_{1k}\,\mu_{1k}^2},
\end{equation}
when the condition $4\,\mu_{1k}^2 - \omega_{k}\,\Omega_{1k}\geq 0$ is fulfilled,
with the remaining critical point $\varrho^c_{k}$ being
\begin{equation}\label{critpoint.1k}
\varrho^c_{k} = \sqrt{\frac{4\,\mu_{1k}^2 - \omega_{k}\,\Omega_{1k}}{4\,\mu_{1k}^2 + \omega_{k}\,\Omega_{1k}}},
\end{equation}
\item $E_{jk}$, when $\Omega_{jk}$ promotes the transition $\omega_j \rightleftharpoons \omega_k$, with $\varrho^c_{k}=\eta_{c}\,\varrho^c_{j}$ in the limit $\varrho^c_{j}\to\infty$,  given by
\begin{equation}\label{eq.Ejpkp}
E_{jk} = \omega_j -\frac{\left((\omega_{k}-\omega_{j}) \,\Omega_{jk }-4\,\mu_{jk}^2\right)^2}{16\,\Omega_{jk}\,\mu_{jk}^2},
\end{equation}
when the condition $4\,\mu_{jk}^2 -(\omega_{k}-\omega_{j}) \,\Omega_{jk}\geq 0$ is fulfilled,
with the remaining critical point $\eta^c_{k}$ being
\begin{equation}\label{critpoint.jk}
\eta^c_{k} = \sqrt{\frac{4\,\mu_{jk}^2 -(\omega_{k}-\omega_{j}) \,\Omega_{jk}}{4\,\mu_{jk}^2 + (\omega_{k}-\omega_{j})\,\Omega_{jk}}},
\end{equation}
\end{itemize}

From the above, the minimum energy that of the variational ground state is given by 
\begin{eqnarray}
E_{min} = \min\left\{E_N,\, \{E_{1k}\},\,\{ E_{j k}\}\right\}\,,
\end{eqnarray}
for $j<k=2,\,3,\,\dots,\,n$.

\subsection{Separatrix and order of transitions}

The separatrix divides the minimum energy surface in regions where the properties of the ground state change significantly. An example of this is the boundary between the normal region, where the ground state is given by the vacuum state for the field and all atoms in their lowest level, and the collective region where the ground state possesses contributions from states with a nonzero number of photons and/or excited atomic states.

For values of the dipolar intensities satisfying
\begin{equation}\label{eq.normalregion}
\begin{array}{l } 4 \,\mu_{1k}^2\leq \omega_{k}\,\Omega_{1k}\,, \\[3mm] 
4\, \mu_{jk}^2\leq (\sqrt{\omega_{j}}+\sqrt{\omega_{k}})^2\,\Omega_{jk}\,,\end{array}
\end{equation}
the minimum energy value is $E_{min}=E_N=0$. The separatrix that divides the normal and collective regions is given by the points where the equality in equations~(\ref{eq.normalregion}) is attained. The separatrices in the collective region are given by the sets of equalities $E_{jk}=E_{j'k'}$, for the modes $\Omega_{jk}\neq \Omega_{j'k'}$:
\begin{eqnarray}
4\, \mu_{jk}^2 -\frac{\Omega_{jk}}{8\,\mu_{j'k'}^2\,\Omega_{j'k'}} \left(\frac{\zeta_{j}+\zeta_{k}}{2} \pm\sqrt{\zeta_{j}\,\zeta_{k}}\right) = 0\,,
\end{eqnarray}
where
\begin{eqnarray}
\zeta_{i} &=& 16\,\mu_{j'k'}^4+\left(\omega_{k'}-\omega_{j'}\right)^2\Omega_{j'k'}^2 \nonumber \\
&+& 8\,\mu_{j'k'}^2\,\Omega_{j'k'}\left(2\,\omega_{i} - \omega_{k'}-\omega_{j'}\right)\,,
\end{eqnarray}
with $i = j,\,k$, and the appropriate sign is taken to lie in the collective regime.

From the expressions for the minimum energy value in the different regions, and following the Ehrenfest classification~\cite{gilmore93}, one finds the order of transitions at each separatrix as
\begin{center}
\begin{tabular}{c |c }
\hspace{5mm}Transition\hspace{5mm} & \hspace{5mm}Order\hspace{5mm} \\[1mm] \hline &\\[-2mm]
$E_N \rightleftharpoons E_{1k}$ & Second  \\[1mm]
$E_N \rightleftharpoons E_{jk}$ & First  \\[1mm]
$E_{jk} \rightleftharpoons E_{j'k'}$ & First\\[1mm]
\end{tabular}
\end{center}
We stress that the critical points where a second order transition occurs form bifurcations, i.e., where the points coalesce at the origin in the $\varrho$-space and hence all physical quantities (expectation value of the number of photons, atomic populations, etc.) change in a continuous manner. The critical points where a first order transition occurs form a Maxwell set, i.e., there are two different critical points with the same energy value. In this case at least one physical quantity has a discontinuity.

\begin{table*}
\begin{center}
\caption{Expectation values and dispersions of the number of photons and atomic populations per particle for the variational ground state in $n$-level systems, in the normal $N$ and collective regions $S_{jk}$  of the phase diagram. We have defined $\phi_{jk}\equiv\phi_{k}-\phi_{j}$, and the condition to have a minimum is $\phi_{jk}=0,\,\pi$. In the text we have taken $\omega_{1}=0$ without loss of  generality.}
\label{tnniv}
\vskip2mm
\begin{tabular}{l | c | c}
 & \ {N} \ &  {$S_{jk}$} \\ \hline\hline & \\[-2mm] 
$\langle \bm{\nu}_{jk}\rangle$ & 0 &  $\displaystyle\frac{\mu_{jk}^2}{\Omega_{jk}^2}\left(1 - \frac{(\omega_k-\omega_j)^2\,\Omega_{jk}^2}{8\,\mu_{jk}^4}\right)$   \\[6mm]
\hline&& \\[1mm]
$\langle \bm{A}_{jj}\rangle$ & $\delta_{j\,1}$ & $\displaystyle\frac{1}{2} \left(1+ \frac{(\omega_k-\omega_{j})\,\Omega_{jk}}{4\,\mu_{jk}^2}\right)$ \\[6mm]

$\langle \bm{A}_{kk}\rangle$ & 0 & $ \displaystyle\frac{1}{2} \left(1- \frac{(\omega_k-\omega_{j})\,\Omega_{jk}}{4\,\mu_{jk}^2}\right)$ \\[6mm]
$\langle \bm{A}_{jk}\rangle$ & 0 & $\displaystyle\frac{1}{2}\left(1-\frac{(\omega_{k}-\omega_{j})^{2}\,\Omega_{jk}^{2}}{16\,\mu_{jk}^{4}}\right)^{1/2}\,e^{i\,\phi_{jk}}$  \\[6mm]
\hline&& \\[1mm]
$(\Delta\bm{A}_{jj})^2$ &0& $\displaystyle\frac{1}{4}\left(1 - \frac{(\omega_k-\omega_j)^2\Omega_{jk}^2}{16\,\mu_{jk}^4}\right)$ \\[6mm]
$(\Delta\bm{A}_{kk})^2$ &0& $\displaystyle\frac{1}{4}\left(1 - \frac{(\omega_k-\omega_j)^2\Omega_{jk}^2}{16\,\mu_{jk}^4}\right)$ \\[6mm]
\end{tabular}
\end{center}
\end{table*}

\subsection{Expectation values} 

The expectation values of matter and field observables for the variational ground state, such as number of photons and atomic population, may be calculated analytically. That for the product of matter and field observables reduces to the product of the expectation values of each one, i.e.,
	\begin{equation}
	\langle\bm{O}_{M}\otimes\bm{O}_{F}\rangle=\langle\bm{O}_{M}\rangle\,\langle\bm{O}_{F}\rangle\ .
	\end{equation}
For this reason, in this variational approximation, there are no correlations between matter and field operators.
In a similar manner as with the minimum energy surface, it is necessary to consider the normal and the different collective regions. In the normal regime one finds that the variational ground state is given by
\begin{equation}
|\psi_N\rangle =  \frac{\bm{b}_1^{\dag\, N_a}}{\sqrt{N_a!}}|0\rangle_M \otimes |0\rangle_F\,,
\end{equation}
where $|0\rangle_M$ and $|0\rangle_F$ denote the vacuum state in the Fock basis for matter and field respectively. Clearly, we have a zero expectation value for the number of photons, and the atomic populations for this state is the corresponding to the lowest atomic level, i.e., $\langle \bm{A}_{11}\rangle = 1$ (per particle). On the other hand, in the collective region, if the minimum energy surface is given by $E_{jk}$ for fixed values of $j$ and $k$ with $j<k=2,\,\dots,\,n$,  the variational ground state takes the form
\begin{eqnarray}
|\Psi_{1k}\rangle &=&  \frac{1}{\sqrt{N_a!}}\left(\frac{\bm{b}_1^\dag + \varrho_k^c\, \bm{b}_k^\dag}{\sqrt{1+{\varrho_k^c}^2}}\right)^{N_a}\,|0\rangle_M\nonumber\\[3mm]
&\otimes&\exp(-N_a\,r_{1k}^{c\, 2}/2)\,\exp(\sqrt{N_a}\,r_{1k}^{c}\,\bm{a}_{1k}^{\dagger})
\,|0\rangle_F\,,\quad
\end{eqnarray}
when $j=1$, with
\begin{eqnarray}
r_{1k}^c = \frac{2\, \mu_{1k}\,\varrho_{k}^c}{(1+{\varrho_k^c}^2)\Omega_{1k}}\,,
\end{eqnarray}
and
\begin{eqnarray}
|\Psi_{jk}\rangle &=&  \frac{1}{\sqrt{N_a!}}\left(\frac{\bm{b}_j^\dag + \eta_k^c\, \bm{b}_k^\dag}{\sqrt{1+{\eta_k^c}^2}}\right)^{N_a}\,|0\rangle_M\nonumber\\[3mm]
&\otimes&\exp(-N_a\,\tilde{r}_{jk}^{c\, 2}/2)\,\exp(\sqrt{N_a}\,\tilde{r}_{jk}^{c}\,\bm{a}_{jk}^{\dagger})
\,|0\rangle_F\,,\quad
\end{eqnarray} 
when $j>1$ and with
\begin{eqnarray}
\tilde r_{jk}^c = \lim_{\varrho_{j}^c\to\infty}\,r_{jk}^c\bigg|_{\varrho_k^c = \eta_k^c\,\varrho_j}=  \frac{2\, \mu_{jk}\,\eta_{k}^c}{(1+{\eta_k^c}^2)\Omega_{jk}}\,.
\end{eqnarray}
The indices $jk$ indicate that we only have photons of mode $\Omega_{jk}$ and that only the atomic levels $j$ and $k$ are populated, with zero contribution from the rest. Table~\ref{tnniv}, shows the different expectation values and fluctuations for the normal and collective regions $S_{jk}$.  One should note that the photon distribution is a Poisson distribution, i.e., one has $(\Delta\,\nu_{jk})^2 = \langle \bm{\nu}_{jk}\rangle$, while the matter contribution is described by a binomial distribution
\begin{eqnarray}
P(x) = \left(\begin{array} {c} N_a \\ x \end{array}\right)\, p^{N_a -x}\, q^x,
\end{eqnarray}
where $p$ stands for the atomic population per particle of the $j$th level  and $q$ for the corresponding one of the $k$th level, {\em i.e.}, from table~\ref{tnniv} one has
\begin{eqnarray}
p &=& \frac{1}{2} \left(1+ \frac{(\omega_k-\omega_{j})\,\Omega_{jk}}{4\,\mu_{jk}^2}\right),\nonumber \\[3mm]
q &=& \frac{1}{2} \left(1- \frac{(\omega_k-\omega_{j})\,\Omega_{jk}}{4\,\mu_{jk}^2}\right), \nonumber
\end{eqnarray} 
and hence the corresponding atomic populations are related to them  by $\langle \bm{A}_{jj}\rangle = N_a\, p$, and $\langle \bm{A}_{kk}\rangle = N_a\, q$.

\subsection{Symmetries in the RWA }

When the RWA approximation is considered the terms that do not preserve the total number of excitations in the Hamiltonian are neglected,  i.e., the Hamiltonian is written as
\begin{eqnarray}\label{eq.H.RWA}
\bm{H}_{\rm \tiny RWA} = \bm{H}_D - \frac{1}{\sqrt{N_a}} \sum_{j<k}^{n} \mu_{jk} \left(\bm{a}_{jk}^\dag\,\bm{A}_{jk}+\bm{a}_{jk}\,\bm{A}_{kj}\right)\,,\qquad
\end{eqnarray}
where $\bm{H}_D$ is given by Eq.~(\ref{eq.HD}). 
Using the same test state discussed previously, one finds the corresponding  variational energy surface in RWA approximation to be
\begin{eqnarray}
{\cal E}_{\rm \tiny  RWA}& = &\sum_{j<k}^n \Omega_{jk}\, {R}_{jk}^2 +  N_a \sum_{j=1}^n\omega_j \frac{\varrho_j^2}{1+\sum_{k=2}^n \varrho_k^2}  \nonumber \\
 && - 2 \sqrt{N_a} \sum_{j<k}^n \mu_{jk}\,{R}_{jk}\frac{\varrho_{j}\,\varrho_{k}\,\cos(\phi_{k}-\phi_{j}-\theta_{jk})}{1+\sum_{k=2}^n \varrho_k^2}.\nonumber
\end{eqnarray}

Comparing with Eq.~(\ref{eq.E.test}) one sees that the effect of considering this approximation modifies the dipolar intensities by a factor of $1/2$; we may therefore find the solutions to the problem in this approximation by replacing $\mu_{jk}\to \mu_{jk}/2$ in all expressions for the critical points of the complete Hamiltonian. Concerning the phases, the critical points are given by the relationship $\phi_{k}^{c}-\phi_{j}^{c}-\theta_{jk}^{c}=0,\,\pi$ with the condition for being a minimum $\mu_{jk}\,\cos(\phi_{k}^{c}-\phi_{j}^{c}-\theta_{jk}^{c})>0$.

On the other hand, one may prove that the Hamiltonian~(\ref{eq.H.RWA}) possesses $n$ linearly independent constants of motion $\bm{K}_j$ (including the total number of particles $N_a$), by using the relationships
\begin{eqnarray}
\bm{a}\,g(\bm{\nu}) &=& g(\bm{\nu}+1)\,\bm{a}\, ,\\[3mm] 
\bm{a}^\dag\,g(\bm{\nu}) &=& g(\bm{\nu}-1)\,\bm{a}^\dag\,,\\[3mm]
\bm{A}_{jk}\,g(\bm{A}_{jj},\bm{A}_{kk}) &=& g(\bm{A}_{jj}-1,\bm{A}_{kk}+1)\,\bm{A}_{jk}\,,\\[3mm]
\bm{A}_{jk}\,g(\bm{A}_{jj},\bm{A}_{kk}) &=& g(\bm{A}_{jj}+1,\bm{A}_{kk}-1)\,\bm{A}_{kj}\,,
\end{eqnarray}
for field $\bm{\nu}=\bm{a}^\dag\,\bm{a}$ and matter $\bm{A}_{jk}=\bm{b}_j^\dag\,\bm{b}_k$ operators,  and where $g(\cdot)$ stands for an arbitrary analytical function. These constants of motion are given in general by
\begin{equation}
\bm{K}_{j} =  \bm{A}_{jj} + \sum_{k<j} \bm{\nu}_{kj} - \sum_{j< k} \bm{\nu}_{jk}\,.
\end{equation} 
Clearly the first order Casimir operator in terms of these constants is given by
\begin{equation}
N_a = \sum_{j=1}^n \bm{K}_j\,.
\end{equation}
Also, the total number of excitations is written as 
\begin{equation}
\bm{M} = \sum_{\ell=2}^n\lambda_\ell\,\bm{K}_\ell = \sum_{j<k}^n\bm{\nu}_{jk} + \sum_{k=2}^n \lambda_k \,\bm{A}_{kk}
\end{equation}
where the integer values of $\lambda_k$, which depend on the particular atomic configuration, stand for the number of excitations that are required  to excite one atom from its lowest atomic level to the $k$th atomic level. As an example of this table~\ref{t.lambdak} shows the values of $\lambda_k$  for the case of  $3$-level atoms in their three atomic configurations.

\begin{table}
\caption{$3$-level atomic configurations: shown are the different values of $\lambda_k$ which correspond to the number of excitations that are required  to excite a single atom from its lowest level to $k$th level. }\label{t.lambdak}
\begin{center}
\begin{tabular}{c|c|c}
\hline
configuration & $\phantom{\displaystyle\frac{1}{2}}\lambda_2\phantom{\displaystyle\frac{1}{2}}$ & $\phantom{\displaystyle\frac{1}{2}}\lambda_3\phantom{\displaystyle\frac{1}{2}}$ \\ \hline
$\phantom{\displaystyle\frac{1}{2}}\Xi\phantom{\displaystyle\frac{1}{2}}$ & 1 & 2\\ 
$\phantom{\displaystyle\frac{1}{2}}V\phantom{\displaystyle\frac{1}{2}}$ & 1 & 1\\
$\phantom{\displaystyle\frac{1}{2}}\Lambda\phantom{\displaystyle\frac{1}{2}}$ & 0 & 1\\
\end{tabular}
\end{center}  
\end{table} 

It is interesting to note that for each constant of motion of the Hamiltonian in the RWA approximation~(\ref{eq.H.RWA}) one has a corresponding symmetry operator for the full Hamiltonian~(\ref{eq.H}). This is  given by
\begin{equation}\label{eq.opPi}
\bm{\Pi}_j = \exp\left(i\,\pi\,\bm{K}_j\right).
\end{equation}
These operators may be useful to define the symmetry-adapted variational states, which provide a better approximation to the exact quantum ground state.

The constants of motion in the RWA approximation will lead to a better variational test function as was shown for the two-level atoms~\cite{castanos09b, castanos11b}. This is done by truncating the power series expansion of the coherent states. For the generalized Dicke case the parity symmetry operators will be useful to define the symmetry-adapted variational states, which provide a better approximation to the exact quantum ground state.

\section{Comparison with the quantum ground state}\label{s.quantum}

In order to compare the above variational solution with the exact quantum one, we consider $3$-level atoms in the $\Xi$-configuration. Using the Fock basis
\begin{equation}
|\nu_{12},\,\nu_{23},\,n_1,\,n_2,\,n_3\rangle\,,
\end{equation} 
for the two modes and the three atomic levels one may diagonalise numerically the Hamiltonian in order to find the ground state. However, one may use the fact that this system  possesses three different symmetries $\bm{\Pi}_j=\exp(i\, \pi\, \bm{K}_j)$~(\ref{eq.opPi}) for $j=1,\,2,\,3$ with
\begin{eqnarray}
\bm{K}_1 &=& \bm{A}_{11} - \bm{\nu}_{12}, \nonumber\\[3mm]
\bm{K}_2 &=& \bm{A}_{22}  + \bm{\nu}_{12} - \bm{\nu}_{23}, \\[3mm]
\bm{K}_3 &=& \bm{A}_{33} + \bm{\nu}_{23}. \nonumber
\end{eqnarray}
Also, both the total number of excitations $\bm{M}_\Xi$ and the number of atoms $N_a$ may be written as a linear combination of them  as $\bm{M}_\Xi = \bm{K}_2 + 2\,\bm{K}_3$ and $N_a = \bm{K}_1+\bm{K}_2+\bm{K}_3$. Using these and $\bm{K}_3$ one may rewrite the Fock basis as
\begin{equation}\label{eq.qket}
|M -N_a -K_3 + n_1,\, K_3-n_3,\, n_1,\, N_a-n_1-n_3,\, n_3\rangle\,,
\end{equation}  
where $M$ and $K_{3}$ are the eigenvalues of  $\bm{M}_\Xi$ and $\bm{K}_3$, respectively.
In fact, using the corresponding symmetry operator for the total number of excitations $\bm{\Pi}_M = \exp(i\,\pi\,\bm{M}_\Xi)$  and the $\bm{\Pi}_3$ operator one finds that the Hamiltonian~(\ref{eq.H}) is divided into four blocks which preserve the parity of the $\bm{M}_\Xi$ and $\bm{K}_3$ operators, {\em i.e.},   the Hilbert space is divided into four ortogonal subspaces 
\begin{eqnarray}
\{|\psi_{ee}\rangle\},\quad \{|\psi_{eo}\rangle\},\quad \{|\psi_{oe}\rangle\},\quad \{|\psi_{oo}\rangle\},
\end{eqnarray}
where the subindexes $e=$ even and $o=$ odd denote the parity of $\bm{M}_\Xi$ and $\bm{K}_3$, respectively. Using these subbasis the Hamiltonian may be diagonalised numerically and hence the minimum energy value $E_{ee},\, E_{eo},\, E_{oe},\, E_{oo}$ for each subspace may be found. Thus, the exact quantum ground energy is given by
\begin{equation}\label{eq.eminq}
E_{min} = \min\{E_{ee},\, E_{eo},\, E_{oe},\, E_{oo}\}\,.
\end{equation} 
%
\begin{figure}
\begin{center}
\includegraphics[width=0.95\linewidth]{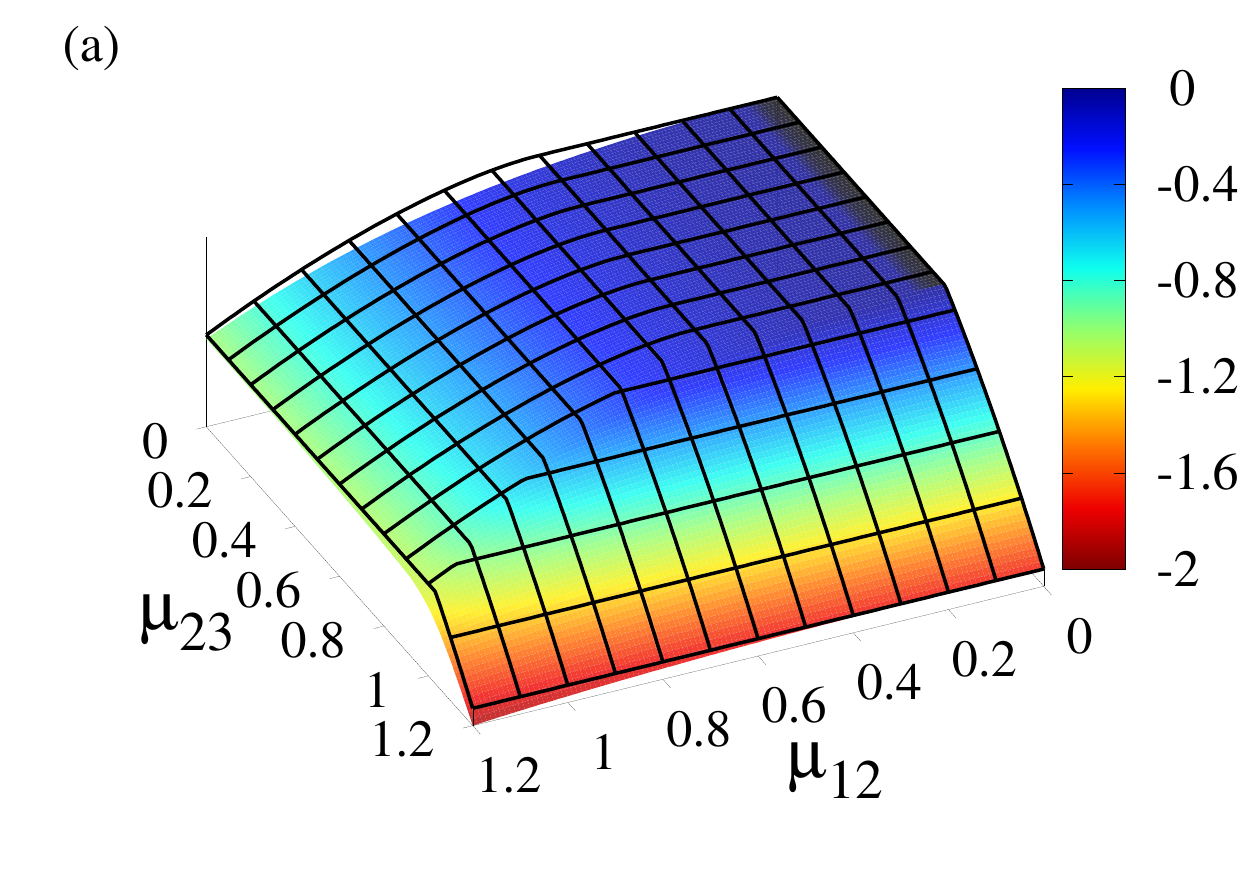}\\[2mm]
\includegraphics[width=0.95\linewidth]{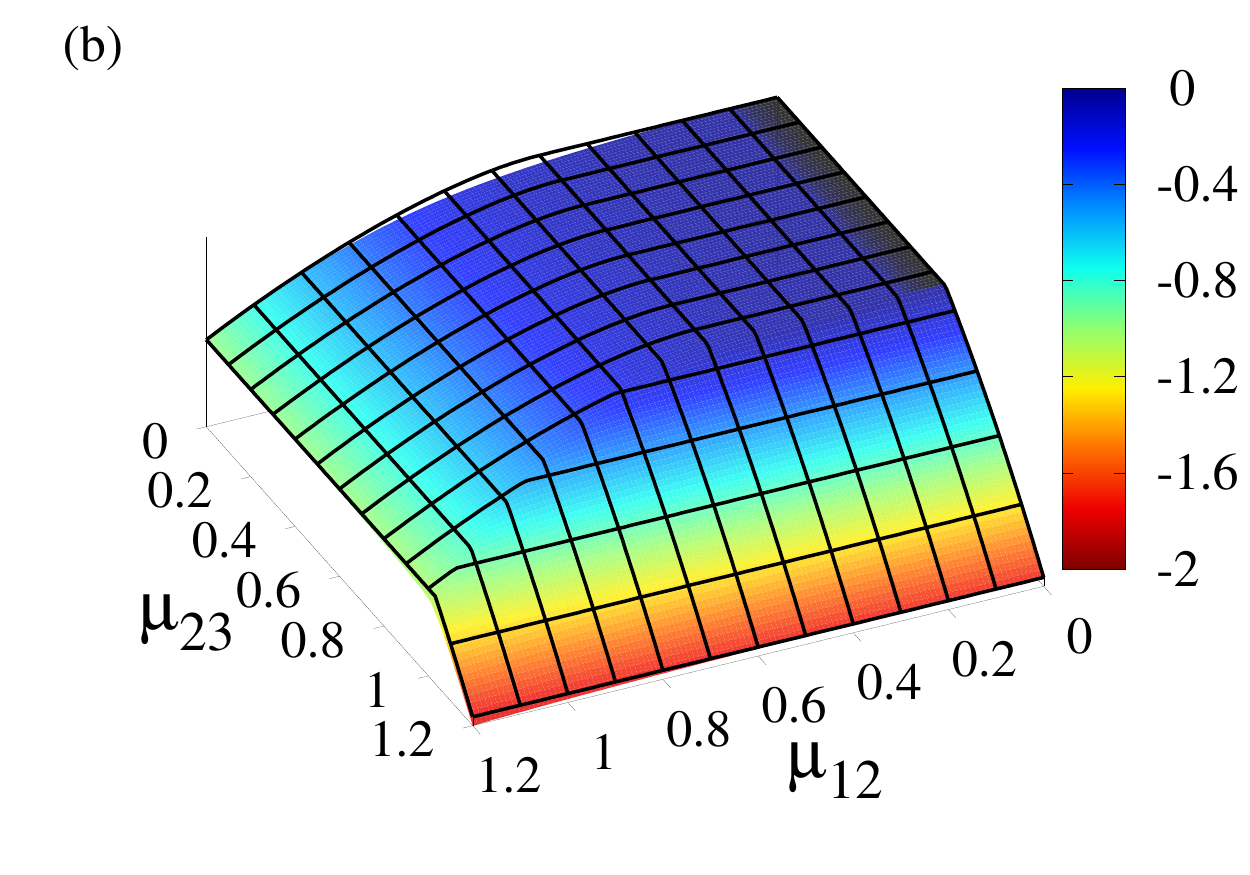}
\end{center}
\caption{The quantum ground energy surface $E_{min}$ (color graded scale) in comparison with the corresponding variational ground energy (mesh surface) as a function of the control parameters for a number of particles (a) $N_a=1$ and (b) $N_a=2$.}\label{eminq}
\end{figure}

Figure~\ref{eminq} shows the quantum ground energy per particle (\ref{eq.eminq}) (color graded scale) in comparison with the corresponding variational ground energy (mesh surface). One observes in both cases $N_a=1$ Fig.~\ref{eminq}(a) and $N_a=2$ Fig.~\ref{eminq}(b) that the variational ground energy surface is  a very good approximation to the exact quantum calculation.

On the other hand, it is of interest to investigate the polychromatic behavior of the phase diagram. For this, we consider the ratio of the difference to the sum of the expectation values of the number of photons, defined by   
\begin{equation}
\delta\, \nu := \frac{\langle \bm{\nu}_{23}-\bm{\nu}_{12}\rangle}{\langle \bm{\nu}_{23}+\bm{\nu}_{12} \rangle}\,,
\end{equation}
which takes values $-1\leq \delta\,\nu \leq 1$ except when the state satisfies $\langle \bm{\nu}_{23}+\bm{\nu}_{12} \rangle=0$, which is satisfied only in the normal region with $\mu_{12}=0$ (black region in figure~\ref{eminq}). 
%
\begin{figure}
\begin{center}
\includegraphics[width=0.95\linewidth]{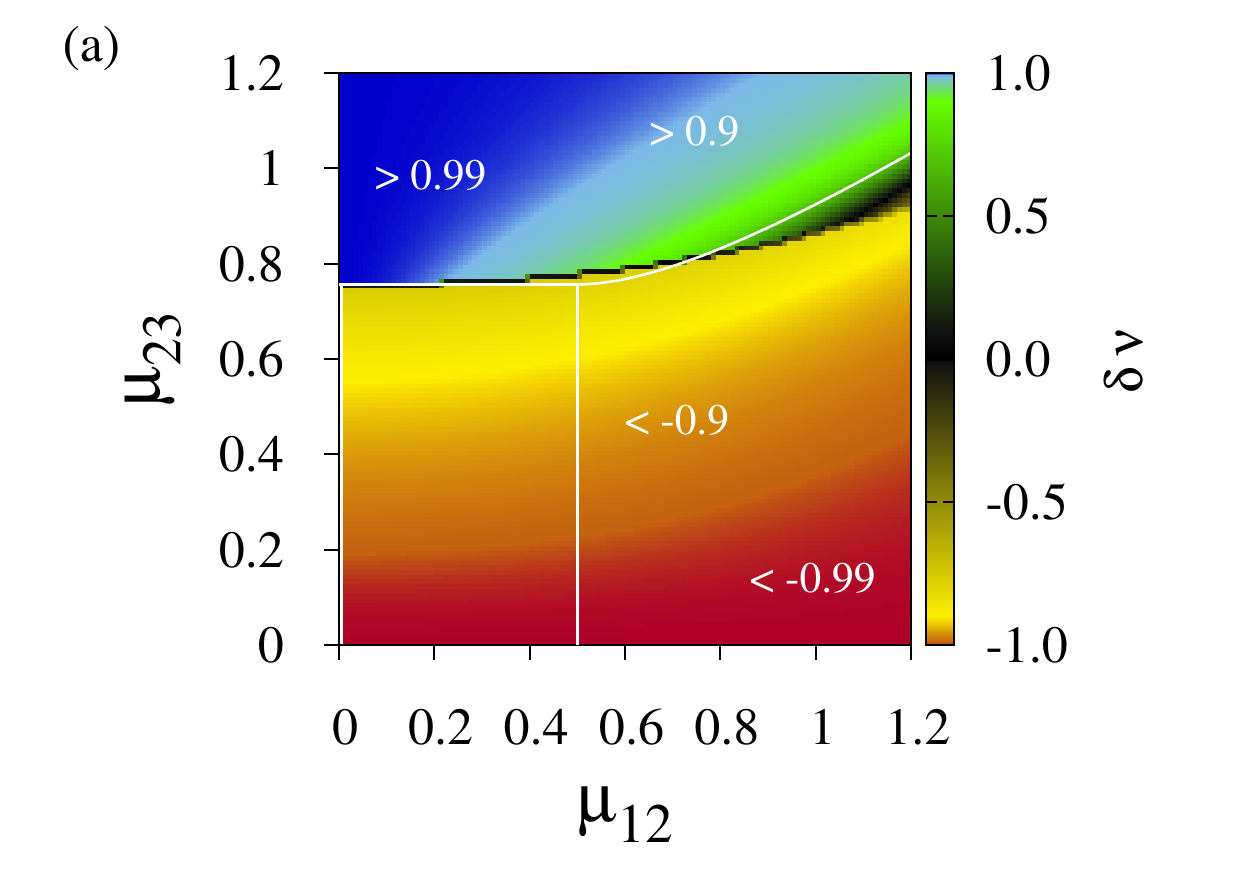}\\[2mm]
\includegraphics[width=0.95\linewidth]{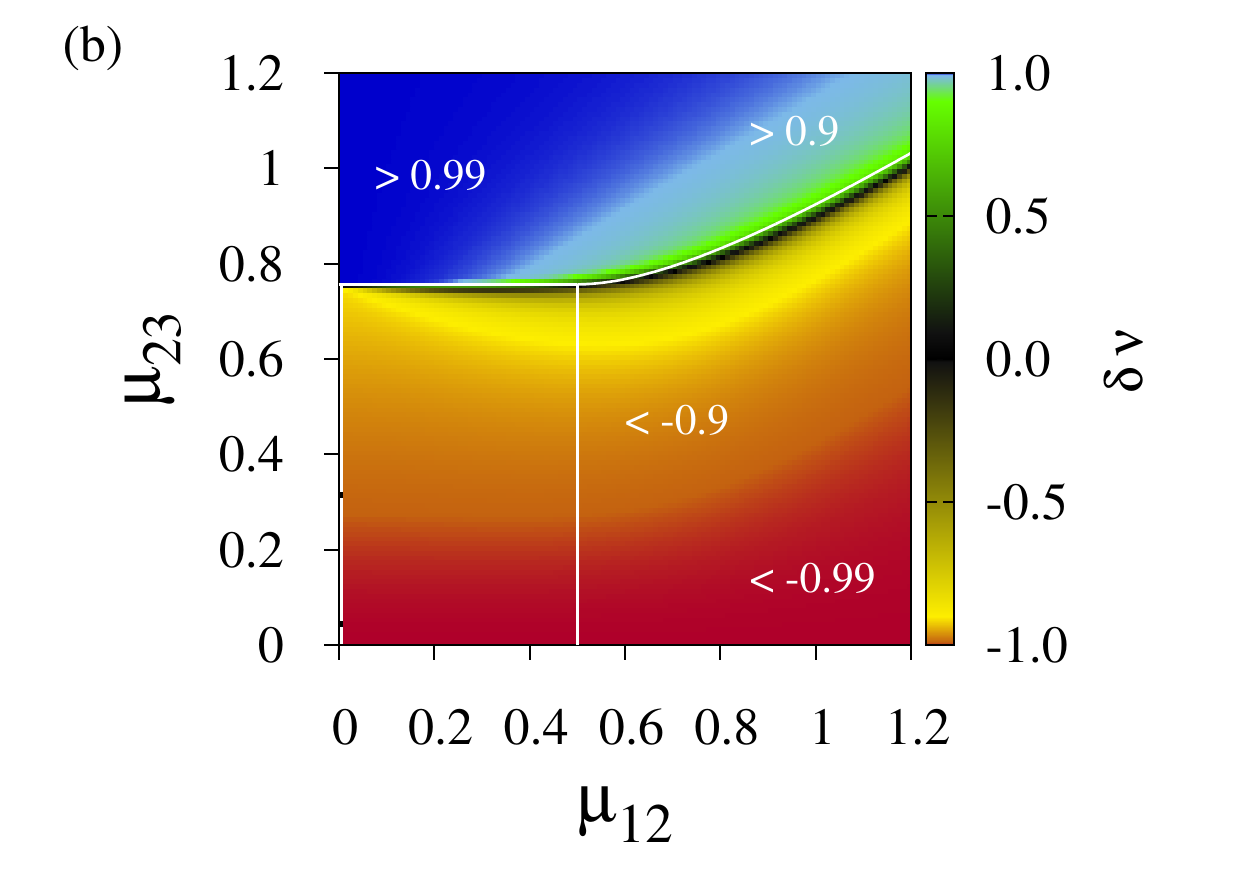}
\end{center}
\caption{The rate of the difference to the sum of the expectation value of the number of photons, $\delta \nu$, of the exact ground state for (a) $N_a=1$ and (b) $N_a=2$.}\label{deltanu}
\end{figure}
Clearly, when $\delta\,\nu \approx -1$ the ground state is dominated by the mode $\Omega_{12}$, in contrast with values $\delta\,\nu \approx 1$ where the state is dominated by the mode $\Omega_{23}$. 

Figure~\ref{deltanu} shows the ratio of the difference to the sum of the expectation values of the number of photons, $\delta\,\nu$, for the corresponding ground states of figure~\ref{eminq}. The classical separatrix is shown in white lines. One may note that this quantity is not sensitive where second order transitions occur (cf. Fig.~\ref{enXVL}\,(a)) and takes values $\delta\,\nu \approx -1$ in both the normal and collective regions, indicating that the mode $\Omega_{12}$ dominates, but around the separatrix where the first order transition occurs both modes $\Omega_{12}$ and $\Omega_{23}$ contribute to the ground state, since $\delta\,\nu \approx 0$. The region above the separatrix is dominated by the mode $\Omega_{23}$ where one has $\delta\,\nu\approx 1$. We note that the quantum separatrix where a first order transition occurs ($\delta \nu =0$, black region in figure~\ref{deltanu}) approaches to the variational separatrix when the number of particles grows, in fact for $N_a=2$ the quantum separatrix is very close to the variational one Fig.~\ref{deltanu}(b), in contrast with the case $N_a=1$  Fig.~\ref{deltanu}(a).  

\begin{figure}
\begin{center}
\includegraphics[width=0.95\linewidth]{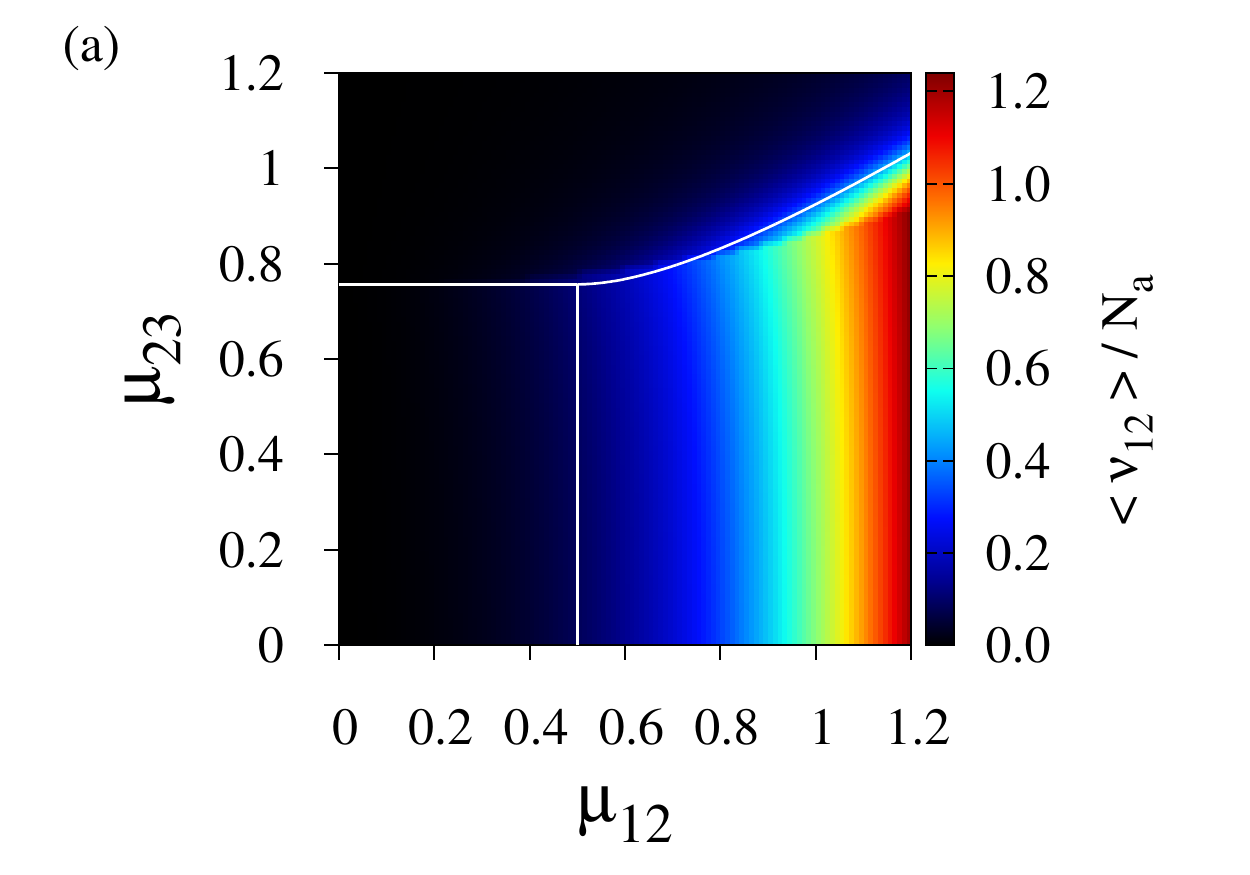}\\[2mm]
\includegraphics[width=0.95\linewidth]{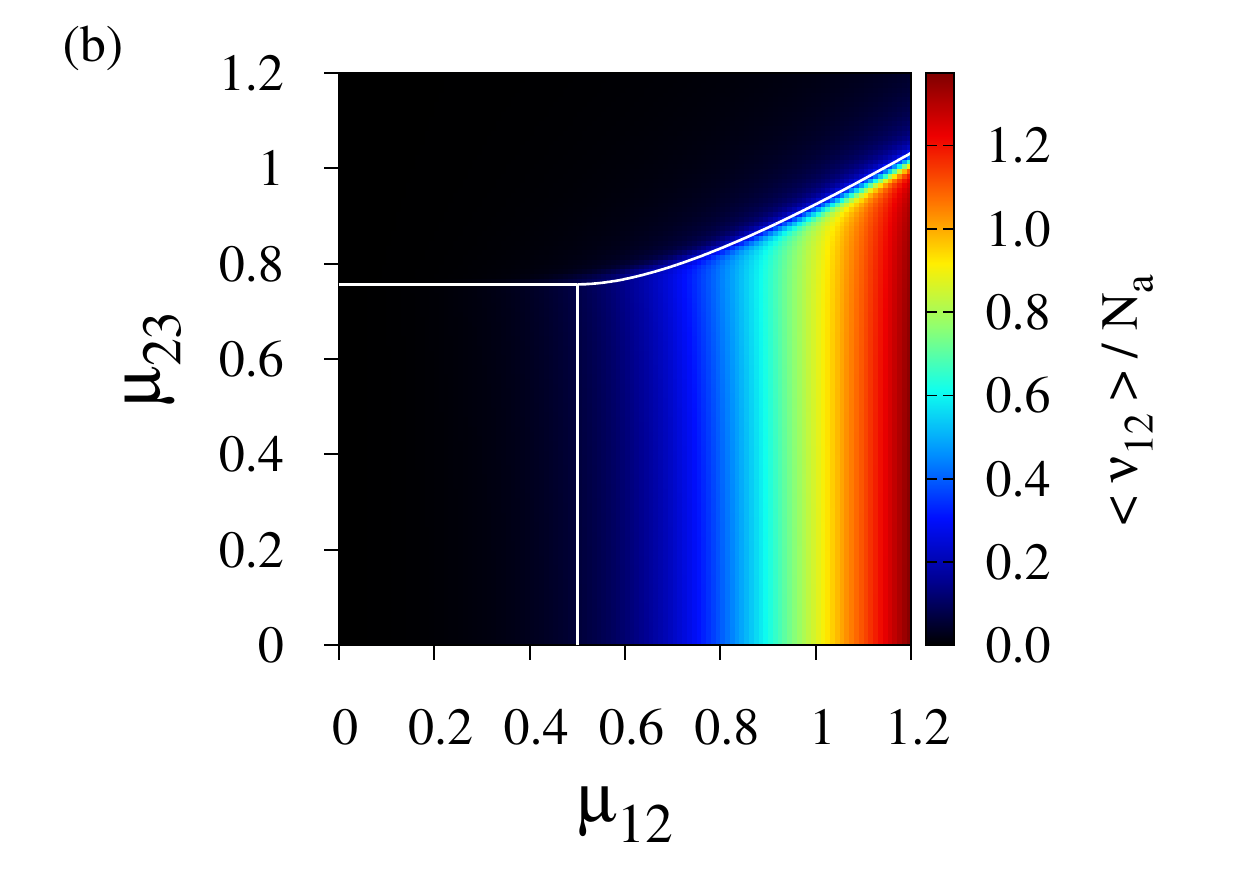}
\end{center}
\caption{Expectation value of the number of photons $\langle \bm{\nu}_{12}\rangle$ of the exact ground state for (a) $N_a=1$ and (b) $N_a=2$. The classical separatrix is indicated (white line).}\label{nu12}
\end{figure}

The expectation value $\langle \nu_{12}\rangle$ is negligible in the normal region; this fact is shown in figure~\ref{nu12}, and we see that the quantum phase diagram is divided into monochromatic regions in a similar fashion to the variational calculation. This is now only in the sense that the bulk of the ground state is dominated by states with only one type of photons in each region; in fact this figure shows that in the normal region the ground state is dominated by the vacuum state.

%
%

\section{Conclusions}\label{conclusions}

A  system of $N_a$ atoms of $n$-levels interacting dipolarly with $\ell$ modes of electromagnetic field, where the transitions between two given atomic levels are promoted only by one mode, has been considered. Using as a variational test state the direct product of Heisenberg-Weyl states for the field contribution, and coherent states preserving the number of atoms for the matter contribution, the variational energy surface is calculated and the minimum obtained. 

We have given an iterative procedure for the calculation of the critical points that reduces any system of $n$-levels to $2$-level systems.

The procedure to find the critical points that minimise the energy surface was exemplified for $3$- and $4$-level atoms interacting with $\ell$ modes ($\ell=2$ for $3$-level atoms and $\ell=3,\,4$ for $4$-level atoms). The normal and collective regions were described analytically and was demonstrated that the collective region is divided into $\ell$ monochromatic regions where only one mode of electromagnetic field contributes to the ground state while the other ones remain in the vacuum state.

Studying the transitions between the different regions in the phase diagram, for $3$-level atoms, we find both first and second order quantum phase transitions (cf. Fig.~\ref{enXVL}). First order transitions are directly related to the fact that at least one physical quantity of the system changes in a discontinuous manner (cf. Figs.~\ref{expectX}(b) and \ref{expectX}(c)), which is related to the fact that the set of critical points in the separatrix forms a Maxwell set.  Second order transitions present a continuous behavior (cf. Fig.~\ref{expectX}(a)), and in this case the critical points form bifurcations. Similar results are obtained for the general case of $n$-level atoms interacting with $\ell$ modes.

This variational study suggests the following relationship between  matter and field observables for the ground state:
	\begin{equation}
		\langle\bm{\nu}_{jk}\rangle=4\,\frac{\mu_{jk}^{2}}{\Omega_{jk}^{2}}\,(
		\Delta\bm{A}_{jj})^{2}
	\end{equation}
which, at the separatrix and considering $\omega_1=0$, implies
	\begin{equation}
		\langle\bm{\nu}_{jk}\rangle=\frac{(\sqrt{\omega_{j}}+\sqrt{\omega_{k}})^{2}}{
		\Omega_{jk}}\,(\Delta\bm{A}_{jj})^{2}\ .
	\end{equation}
Being this a universal relationship, one may propose it as an experimental criterion to detect the transition between the normal and superradiant regimes.

We have shown that, when the RWA approximation is considered, the phase diagram of the system suffers a rescaling of the dipolar intensities by replacing $\mu\to (\mu)^{RWA}/2$ in all quantities.  Also, we have shown that in the RWA approximation the Hamiltonian possesses $n$ linearly independent constants of motion (including the total number of excitations), and that each one of them provides a symmetry operator of the full Hamiltonian. These will be useful to establish symmetry-adapted variational states. 

Finally, using the symmetries of the full Hamiltonian we found by numerical calculation  the exact quantum solution for the case of $3$-level atoms in the $\Xi$-configuration. We also found that the variational minimum energy surface is a very good approximation to  the exact quantum one. Also, we considered the ratio of the difference to the sum of the number of photons and found that the quantum phase diagram is divided into monochromatic regions, this in the sense that the bulk of the ground state contains contributions of only one type of photons in each region.

\section*{Acknowledgments}
This work was partially supported by CONACyT-M\'exico (under project 238494), and DGAPA-UNAM (under projects IN101614 and IN110114).

\end{document}